\documentclass[fleqn,usenatbib]{mnras}

\usepackage{newtxtext,newtxmath}

\usepackage[T1]{fontenc}
\usepackage{ae,aecompl}

\usepackage{graphicx}	
\usepackage{subcaption}
\usepackage{amsmath}	
\usepackage[dvipsnames]{xcolor}

\def\JHD#1{\textcolor{blue}{#1}}

\title[MG with WL peaks]{Constraining modified gravity with weak lensing peaks}

\author[C. T. Davies et. al]
{Christopher T. Davies$^{1}$,\thanks{E-mail:C.Davies@physik.uni-muenchen.de (CTD)}
Joachim Harnois-D\'eraps$^{2}$,
Baojiu Li$^{3}$,
Benjamin Giblin$^{4}$,\newauthor
C\'esar Hern\'andez-Aguayo$^{5, 6}$,
and Enrique Paillas$^{7, 8}$
\\
$^{1}$University Observatory, Faculty of Physics, Ludwig-Maximilians-Universit{\"a}t, Scheinerstr. 1, 81679, Munich, Germany\\
$^{2}$School of Mathematics, Statistics and Physics, Newcastle University, Herschel Building, NE1 7RU, Newcastle-upon-Tyne, UK\\
$^{3}$Institute for Computational Cosmology, Department of Physics, Durham University, South Road, Durham DH1 3LE, UK\\
$^{4}$Scottish Universities Physics Alliance, Institute for Astronomy, University of Edinburgh, Blackford Hill, Scotland, UK\\
$^{5}$Max-Planck-Institut f\"ur Astrophysik, Karl-Schwarzschild-Str. 1, D-85748, Garching, Germany\\%
$^{6}$Excellence Cluster ORIGINS, Boltzmannstrasse 2, D-85748 Garching, Germany\\%
$^{7}$Waterloo Centre for Astrophysics, University of Waterloo, Waterloo, ON N2L 3G1, Canada\\
$^{8}$Department of Physics and Astronomy, University of Waterloo, Waterloo, ON N2L 3G1, Canada\\
}

\date{Accepted XXX. Received YYY; in original form ZZZ}

\pubyear{2015}

\begin{document}
\label{firstpage}
\pagerange{\pageref{firstpage}--\pageref{lastpage}}
\maketitle

\begin{abstract}
It is well established that maximizing the information extracted from upcoming and ongoing stage-IV weak-lensing surveys requires higher-order summary statistics that complement the standard two-point statistics. In this work, we focus on weak-lensing peak statistics to test two popular modified gravity models, $f(R)$ and nDGP, using the FORGE and BRIDGE weak-lensing simulations, respectively. From these simulations we measure the peak statistics as a function of both cosmological and modified gravity parameters simultaneously. Our findings indicate that the peak abundance is sensitive to the strength of modified gravity, while the peak two-point correlation function is sensitive to the nature of the screening mechanism in a modified gravity model. We combine these simulated statistics with a Gaussian Process Regression emulator and a Gaussian likelihood to generate stage-IV forecast posterior distributions for the modified gravity models. We demonstrate that, assuming small scales can be correctly modelled, peak statistics can be used to distinguish GR from $f(R)$ and nDGP models at the two-sigma level with a stage-IV survey area of $300 \, \rm{deg}^2$ and $1000 \, \rm{deg}^2$, respectively. Finally, we show that peak statistics can constrain $\log_{10}\left(|f_{R0}|\right) = -6$ to 2\% precision, and  $\log_{10}(H_0 r_c) = 0.5$ to 25\% precision.

\end{abstract}




\section{Introduction}

Understanding the nature of the late-time accelerated expansion of the Universe is one of the primary goals in cosmology.  Whilst the origin of this expansion is presently unknown, it is best fit by a cosmological constant, denoted $\Lambda$. Simultaneously, there are open questions on the nature of cold dark matter (CDM), a non-baryonic and non-relativistic contributor to the mass energy density of the Universe, which is required to explain the mismatch between the apparent strength of gravity on large scales and the relative scarcity of baryonic material. These mysterious components lend their name to the standard model of cosmology, $\Lambda$CDM, on account of their dominance at late times, making up $\sim$96\% of the energy density of the universe.

Recent studies, however, have cast doubt on whether a $\Lambda$CDM cosmological model can simultaneously describe early- and late-time astronomical observations. Constraints on the Hubble parameter, $H_0$, for example, from the temperature and polarisation spectra of the cosmic microwave background \citep[CMB;][]{Planck2018} and from supernovae-/Cepheid-based distance-redshift relations, are in disagreement at the level of 4--6$\sigma$ \citep{verde/etal:2019, riess/etal:2020, riess/etal:2021, di-valentino/etal:2021}. There is also the so-called `$S_8$ tension', referring to the 2--3$\sigma$ difference in low-redshift probes such as weak gravitational lensing \citep[WL;][]{Asgari2020, amon/etal:2021, secco/etal:2021, li/etal:2023, dalal/etal:2023, des/kids:2023} and CMB-derived estimates of $S_8=\sigma_8 \sqrt{\Omega_{\rm m}/0.3}$, where $\sigma_8$ is the amplitude of the matter power spectrum and $\Omega_{\rm m}$ is the matter energy density parameter, with CMB-derived estimates giving higher values of $S_8$. We note, however, that the low redshift cluster abundance constraints from eROSITA \citep{ghirardini2024} report $S_8$ values that are roughly $1.4\sigma$ higher than that from the CMB \citep{Planck2018} 

These open questions in cosmology incentivise exploration of models beyond the $\Lambda$CDM paradigm, as well as investment in alternative cosmological statistics which may yield improvements in constraining power and increased accuracy in the determination of the aforementioned tensions. This paper addresses both of these points, by investigating the capacity of an alternative weak lensing statistic for constraining modified gravity signatures in upcoming data sets.    

WL is a powerful cosmological probe which relies on the small distortions observed in the shapes of galaxies out to redshifts of $\sim 1.5$. These distortions, or `shear', are a consequence of deflections in the trajectories of the galaxies' light caused by the gravity of the intervening large-scale structure. Cosmic shear analyses typically measure the correlations in the shapes and orientations of $>10^8$ galaxies \citep{giblin/etal:2021, gatti/etal:2021b} as a function of their angular separations and redshifts, the strength of which is particularly sensitive to $S_8$ \citep{Bartelmann2001}. Stage-III WL experiments, the Kilo-Degree Survey \citep[KiDS;][]{kuijken/etal:2015}, the Dark Energy Survey \citep[DES;][]{jarvis/etal:2016}, and the Hyper-Suprime Cam consortium \citep[HSC;][]{aihara/etal:2018} have succeeded in measuring $S_8$ with precisions at the $\sim 4\%$ level \citep{ li/etal:2023, dalal/etal:2023, des/kids:2023}, and have paved the way for the upcoming lensing data sets from Stage-IV surveys \citep[e.g.][]{LSST2012, Amendola2013}.

Nevertheless, improvements in the precision of cosmological WL constraints are possible given the convention within the field for summarising data using two-point (2pt) statistics, such as power spectra or their real-space counterparts, correlation functions. This approach is robust but sub-optimal owing to the fact that 2pt-probes extract information only up to second order, whereas the non-Gaussianity of the large-scale structure in our Universe means that there is higher-order information which is only accessible, in principle, to so-called ``beyond-2pt" lensing statistics. These include for example the three-/four-point statistics \citep{takada/jain:2002,kilbinger/schneider:2005,fu/etal:2014,gatti/etal:2021,heydenreich/etal:2022}, one-point density distributions \citep{bernadeau/etal:2000, liu/madhavacheril:2019, thiele/etal:2020, boyle/etal:2021, uhlemann/etal:2022, giblin/etal:2023}, density peak abundance and clustering \citep{kratochvil/etal:2010, J.Liu2015, Shan2018, Martinet2018, harnois2021, davies/etal:2021}, lensing voids \citep{Davies2018, Davies2019b, Davies2020b}, higher-order moments \citep{wang/etal:2009, Petri2013, Petri2016, vicinanza/etal:2016, heydenreich/etal:2021, gatti/etal:2021, gatti/etal:2023}, Minkowski functionals \citep{kratochvil/etal:2012, Petri2013, petri/etal:2015, grewal/etal:2022}, density-split statistics \citep{Gruen2018, friedrich/etal:2018, brouwer/etal:2018, burger/etal:2022}, clipped lensing statistics \citep{Giblin2018}, and convolutional neural networks \citep{fluri/etal:2018, fluri/etal:2019, fluri/etal:2022}.

Many of the aforementioned beyond-2pt lensing statistics are critically dependent on numerical simulations to model their cosmological dependence, covariance, sensitivity to systematics, and the signal on small scales. This fact has motivated the development of large simulation suites, including the F-of-R Gravity Emulator \citep[FORGE;][]{Arnold:2022} and BRaneworld-Inspired DGP Gravity Emulator \citep[BRIDGE;][]{Harnois-Deraps:2022bie} simulations employed in this work to quantify the cosmological constraining power and sensitivity of WL peaks to these two prominent modified gravity (MG) models, $f(R)$ gravity and the normal branch of the Dvali-Gabadadze-Porratti (nDGP) model. 

In general, MG models introduce a fifth force that is of comparable strength to Newtonian gravity. However, such strong deviations from General Relativity (GR) are ruled out by tests from the solar system \citep{Will_2006,Will2014}. This means that screening mechanisms must be present in a viable MG model, which suppress the fifth force in certain regimes (such as within our solar system) and allows for deviations from GR in other regimes. An example of this behaviour can be seen in $f(R)$ gravity, where the fifth force is only unscreened in extensive regions with low density. Generally the parameters in the MG models that determine the extent of the deviation from structure formation in GR represent the prevalence of the screening mechanism throughout the Universe. This is exemplified in \cite{Mitchell_2018}, which provides a fitting function for the masses at which haloes are fully screened as a function of the $f(R)$ gravity parameter. This shows that for the weakest ($\log_{10}\left(|f_{R0}|\right)=-6$) and strongest ($\log_{10}\left(|f_{R0}|\right)=-4.5$) $f(R)$ models considered in this work, haloes with masses above, respectively, $\approx 10^{12.5} M_\odot\ $ and $\approx 10^{14.5} M_\odot\ $, are fully screened at redshift zero. We refer the readers to \cite{Harnois-Deraps:2022bie} for an illustration of how the MG models studied in this work alter the convergence power spectrum relative to GR. Additionally, we note that, for $f(R)$ gravity, GR is recovered when $f_{R0} = 0$ (so more negative values of $\log_{10}\left(|f_{R0}|\right)$ are closer to GR), and for nDGP we recover GR at $H_0r_c = \infty$ (where larger values are closer to GR).

Recent constraints on MG from 2pt WL statistics include \cite{ Harnois_Deraps_2015, Hojjati_2016, Joudaki_2017, Abbott_2019,Troster_2021, Abbott_2023} and tighter constraints using WL peaks include \cite{X.Liu2016}, with constraints on the $f(R)$ parameter $\log_{10}\left(|f_{R0}|\right) < -4.82$. Similar constraints have been achieved with galaxy clusters \citep{artis2024}, which finds an upper limit of $\log_{10}\left(|f_{R0}|\right) < -4.31$, and \cite{Vogt2024b} have shown that when combined with CMB data, galaxy clusters can constrain the $f(R)$ model to $\log_{\rm{10}}(|f_{R0}|) < -5.32$.

In this work we provide a detailed investigation of the response from WL peaks to MG and present the information content of WL peaks for MG in the context of stage-IV WL survey posterior distributions. We note that we sample a wide range of MG strengths in this work, which includes strong models that are already ruled out by \cite{X.Liu2016}, down to much weaker models than those that are ruled out by \cite{X.Liu2016}. As an example, models such as $\log(|f_{R0}|) = -6$ (and similar nDGP model) are of particular interest in this work. This is because MG models of this strength are presently viable given existing constraints, as cosmological tests of MG have not yet reached this level of precision. We will show later the WL peak statistics studied here are able to distinguish between $\log(|f_{R0}|) = -6$ and GR.

As a side note, the $f(R)$ gravity model we study here does not modify the background expansion history, so it should be irrelevant to the $H_0$ tension \citep{Hu:2023}. For the $S_8$ tension, MG may play a role as the model does not affect the CMB but can affect $\sigma_8$ at late times. However, the models we test in this work are unlikely to resolve the $S_8$ tension \citep{Kazantzidis_2021}, as they enhance structure formation and hence increase $S_8$ relative to $\Lambda$CDM. Nonetheless these models provide a testbed for beyond-$\Lambda$CDM structure formation, which is crucial for tests of new physics and the late-time accelerated expansion.

This paper is structured as follows. We introduce the relevant WL and MG theory in Sec.~\ref{sec:theory}, followed by a description of the simulations we employ in this work in Sec.~\ref{sec:numerics}. In Sec.~\ref{sec:methodology} we outline our methodology for investigating the power of WL peaks for differentiating between MG theories in mock data, and we present our results in Sec.~\ref{sect:results}. These are discussed alongside our conclusions in Sec.~\ref{sec:conclusion}.

\section{Theory} \label{sec:theory}
First, we outline the relevant WL theory employed in this work in Sec.~\ref{sec:wl theory}. This is followed by a presentation of the modified gravity models studied in the work, $f(R)$ and nDGP, in Sec.~\ref{subsec:MG}.

\subsection{WL}\label{sec:wl theory} 

Weak lensing causes distortions in the image of background galaxies through convergence, shear, and deflections of light rays.
Under the Born approximation and neglecting lens-lens coupling and other second-order effects, the deflection angle can be ignored, leading to straight line  geodesics for which cosmic distortions can be expressed by a distortion matrix \pmb{A}:
\begin{equation}
    \pmb{A} = 
    \begin{pmatrix}
    1 - \kappa -\gamma_1 & -\gamma_2\\
    -\gamma_2 & 1-\kappa+\gamma_1
    \end{pmatrix}
    \, ,
\end{equation}    
where $\kappa$ and  $\gamma = \gamma_1 + i\gamma_2$, are respectively the convergence and shear imparted by the foreground large-scale structures. Both of these are related to the lensing potential via 
\begin{equation}
    \kappa \equiv \frac{1}{2} \nabla^2_{\pmb{\theta}} \psi \, ,
    \label{eq:convergence}
\end{equation}
and
\begin{equation}
    \gamma_1 \equiv \frac{1}{2}\left(\nabla^2_{\pmb{\theta}_1}-\nabla^2_{\pmb{\theta}_2}\right)\psi, 
    \quad\quad\quad
    \gamma_2 \equiv \nabla_{\pmb{\theta}_1}\nabla_{\pmb{\theta}_2}\psi,
    \label{eq:shear}
\end{equation}
where $\nabla_{\pmb{\theta}} \equiv \chi^{-1}\nabla_{\rm 2D}$, with $\chi$ the comoving distance to the lens. 

Note that the relationship between the Newtonian potential $\Phi$ 
and the non-relativistic matter density contrast, depends explicitly on the gravity model under consideration. In addition, the potential that governs photon trajectories may not be strictly identical to that which governs matter within a given model. For the MG models considered in this work, whilst the matter distribution evolves under a modified Poisson equation (see Sec.~\ref{subsec:MG}), the photon geodesics follow the usual GR Poisson equation:
\begin{equation}
    \nabla^2\Phi = 4 \pi G a^2 \bar{\rho}_{\rm{m}} \delta \, ,
    \label{eq:Poisson equation}
\end{equation}
where $G$ is the gravitational constant, $a$ is the scale factor, $\rho_{\rm{m}}$ is the non-relativistic matter density of the universe (with a bar denoting the cosmic mean) and $\delta\equiv\rho_{\rm{m}}/\bar{\rho}_{\rm{m}} - 1$ is the matter density contrast. For source galaxies located at $\chi$, the WL convergence is given by:
\begin{equation}
    \kappa(\pmb{\theta},\chi) = \frac{3H_0^2\Omega_{\rm{m}}}{2c^2}\int_0^{\chi}\frac{\chi - \chi'}{\chi} \chi' \frac{\delta(\chi'\pmb{\theta},\chi')}{a(\chi')} {\rm d}\chi',
    \label{eq:conv source}
\end{equation}
where $H_0$ is Hubble's constant, $c$ is the speed of light and $\Omega_{\rm m}$ the matter density parameter as defined in the introduction.

The above assumes a fixed source plane, but real observations contain galaxies that spans over a range of $\chi$ values, hence Eq. \eqref{eq:conv source} must be weighted by the full source galaxy distribution $n(\chi)$ in order to obtain $\kappa(\pmb{\theta})$ \citep[see, e.g.,][for details]{Kilbinger2015}
\begin{equation}
    \kappa(\pmb{\theta}) = \int_0^{\chi_{\rm H}} n(\chi') \kappa(\pmb{\theta},\chi') {\rm d}\chi' \, ,
\end{equation}
where $\chi_{\rm H}$ is the comoving distance to the horizon. As discussed in Sec.~\ref{sec:numerics}, the $n(\chi')$ distribution of source catalogues can be further split into redshift bins, allowing for a tomographic analysis of the foreground matter distribution.

WL observations rely on accurately measuring the shapes of galaxies, and cross correlating the shapes of neighbouring galaxies. However, the convergence signal induced by correlations in shapes due to lensing is small compared to the random fluctuations in the shapes and orientations of the source galaxies. This is the leading source of noise in WL observations, referred to as galaxy shape noise (GSN).  

In order to account for this, we generate mock GSN maps by assigning to pixels in the WL field random convergence values from a Gaussian distribution with standard deviation 
\begin{equation}
        \sigma_{\rm{pix}}^2 = \frac{\sigma_{\rm{int}}^2}{2 \theta_{\rm{pix}}^2 n_{\rm{gal}} }
    \;,
    \label{eq: GSN gaussian}
\end{equation}
where $\theta_{\rm{pix}}$ is the width of each pixel, $\sigma_{\rm{int}}$ is the intrinsic ellipticity dispersion of the source galaxies, and $n_{\rm{gal}}$ is the measured source galaxy number density. In this work we use $\sigma_{\rm{int}} = 0.26$ and $n_{\rm{gal}} = 5$ gal arcmin$^{-2}$ per tomographic bin, which gives $n_{\rm{gal}} = 25$ gal arcmin$^{-2}$ when using the full $n(z)$ distribution.

WL peaks are closely related to the dark matter haloes along the line of sight \citep{Yang2011,J.Liu2016, Wei2018, Sabyr2022}. Both the abundance and large-scale clustering of haloes encode useful information about the underlying cosmological model. Therefore, as well as studying the abundance of WL peaks, we will also study their clustering, which was shown to contain useful cosmological information for the $w$CDM model in \cite{davies/etal:2021}. The extent to which objects are clustered can be measured through the two-point correlation function (2PCF) which is defined as the excess probability, relative to a random distribution, of finding a pair of objects at a given separation $\theta$. Formally, this is written as
\begin{equation}
    {\rm d}P_{ij}(\theta) = \bar{n}^2(1+\xi(\theta)){\rm d}A_i {\rm d}A_j \, ,
\end{equation}
where $\bar{n}$ is the mean number density of tracers, ${\rm d}A_i$ and ${\rm d}A_j$ are two area elements separated by a displacement $\boldsymbol{\theta}$, and $\xi(\theta)$ is the 2PCF. For irregular sky coverage, the 2PCF can be measured through the Landy-Szalay estimator \citep{Landy1993}. This estimator accounts for the presence of survey masks, which require the generation of random catalogues and is given by
\begin{equation}
    \xi_{\rm{LS}}(\theta) = 1 + \bigg(\frac{N_R}{N_D}\bigg)^2 \frac{DD(\theta)}{RR(\theta)} - \bigg(\frac{N_R}{N_D}\bigg) \frac{DR(\theta)}{RR(\theta)}.
    \label{Eq: LS estimator}
\end{equation}
Here, $N_D$ and $N_R$ are the numbers of data and random points, and $DD$, $DR$ and $RR$ are the numbers of data-data, data-random and random-random pairs in bins $\theta \pm \delta \theta$, respectively. We refer the reader to \citet{Davies2019} for more details about the measurement of the peak 2PCF, which are important for small lensing maps.

\subsection{Modified gravity models}
\label{subsec:MG}

A large number of models describing modifications to the theory of GR can be found in the literature and have been tested against astronomical and cosmological data. Such models can be used as a testbed to constrain possible deviations from GR based on cosmological observations \citep[see e.g.,][for recent reviews]{Koyama:2015vza,Hou:2023kfp}. 

In many cases, the deviation from the GR or Newtonian law of gravity is described by a fifth force mediated by a new dynamical degree of freedom, such as a scalar field, $\varphi$. This modifies the usual Poisson equation, Eq.~\eqref{eq:Poisson equation}, into the following form:
\begin{equation}\label{eq:poisson_MG}
\boldsymbol{\nabla}^2 \Phi = 4\pi G a^2 \bar{\rho}_{\rm m}\delta + \frac{1}{2}\boldsymbol{\nabla}^2\varphi\,,
\end{equation}
where it is made clear that $\varphi$ -- or actually its spatial perturbation -- acts as the potential of the fifth force. 

The fifth force, if acting at its full strength, can be inconsistent with current tests from laboratory or Solar System experiments. As a result, viable MG models often have some screening mechanism which can suppress the fifth force in regions where such experiments are carried out. The detailed working of the screening mechanism depends on the dynamics of the scalar field, governed by its equation of motion. In one of the MG models that are considered here, known as the normal-branch Dvali-Gabadadze-Poratti model \citep[][nDGP hereafter]{Dvali:2000hr}, the equation of motion is given by:
\begin{equation}\label{eq:phi_dgp}
\boldsymbol{\nabla}^2 \varphi + \frac{r_{\rm c}^2}{3\beta\,a^2c^2} \left[ (\boldsymbol{\nabla}^2\varphi)^2
- \boldsymbol{\nabla}^i\boldsymbol{\nabla}^j\varphi\boldsymbol{\nabla}_i\boldsymbol{\nabla}_j\varphi \right] = \frac{8\pi G a^2}{3\beta} \bar{\rho}_{\rm m}\delta\,,
\end{equation}
in which $r_{\rm c}$ is a model parameter of the dimension of length, $i,j=1,2,3$ and
\begin{equation}\label{eq:beta_dgp}
\beta(a) = 1 + \frac{\Omega_{\rm m}a^{-3} + 2\Omega_\Lambda}{2\sqrt{\Omega_{\rm rc}(\Omega_{\rm m}a^{-3} + \Omega_{\Lambda})}}\,,
\end{equation}
is a time-dependent positive-definite function, with $\Omega_\Lambda \equiv 1 - \Omega_{\rm m}$ and $\Omega_{\rm rc} \equiv 1/(4H^2_0r^2_{\rm c})$. The nDGP model originates from theories of braneworld, in which our Universe is a 3-dimensional brane that is embedded in a 4-dimensional space, with gravity being the only of the fundamental forces that is not confined on the 3-dimensional brane. The way screening works in this model is due to the $(\boldsymbol{\nabla}^2\varphi)^2$ and $\boldsymbol{\nabla}^i\boldsymbol{\nabla}^j\varphi\boldsymbol{\nabla}_i\boldsymbol{\nabla}_j\varphi$ terms in Eq.~\eqref{eq:phi_dgp}, which vastly dominate over the $\boldsymbol{\nabla}^2\varphi$ term in that equation if the density perturbation is large ($\delta\gg1$) and hence help to make $\boldsymbol{\nabla}^2\varphi$ much smaller than the Newtonian term $4\pi G a^2 \bar{\rho}_{\rm m}\delta$ (assuming that $\beta(a)$ is order unity or larger) in Eq.~\eqref{eq:poisson_MG}. This is known as the Vainshtein screening mechanism \citep{Vainshtein:1972plb}. 

In the second class of MG models considered here, a different screening mechanism, called chameleons \citep{Khoury:2003rn,Khoury:2003aq,Brax:2004px,Mota:2006fz}, is employed. One concrete example of chameleon-type screened MG models is $f(R)$ gravity, where the Einstein-Hilbert action is modified to have an additional function of the Ricci scalar $R$, $f(R)$ \citep{Buchdahl:1970},
\begin{equation}\label{eq:S-f(R)}
S = \frac{1}{16\pi G} \int \mathrm{d}^4x\sqrt{-g} \left(R + f(R)\right)\,,
\end{equation} 
where $g_{\mu\nu}$ is the metric tensor, $g \equiv \rm{det}(g_{\mu\nu})$ and the action of the matter fields are omitted as they are standard. This addition, while simple, leads to substantial changes to the gravitational field equations, notably by making it contain up to fourth-order derivatives of the metric field. Nevertheless, this model can be recast as standard GR non-minimally coupled to a scalar field $\varphi = -f_R \equiv {\rm d} f(R) / {\rm d} R$. The equation of motion for the field $f_R$ is given, in the weak-field limit, by 
\begin{equation}\label{eq:fR}
\boldsymbol{\nabla}^2 f_R = -\frac{a^2}{3} \left[R(f_R) - \bar{R} + 8\pi G\bar{\rho}_{\rm m}\delta\right]\,,
\end{equation}
where $R$ is now expressed as a function of $f_R$ by reverting $f_R(R)$ and $\bar{R}$ is the cosmic mean of the Ricci scalar. It is evident from here that the detailed behaviour of the model depends on the functional form of $f(R)$, though different choices of $f(R)$ -- as long as they guarantee chameleon screening -- often only give quantitative differences in their predictions. In this work we consider the \citet{Hu:2007nk} $f(R)$ model, with
\begin{equation}\label{eq:f(R)}
f(R) = -m^2 \frac{c_1}{c_2} \frac{(-R/m^2)^n}{\left(-R/m^2\right)^n + 1}\,,
\end{equation}
where $m^2 \equiv \Omega_{\rm m} H_0^2$, and $c_1, c_2$ and $n$ are free dimensionless model parameters, with $n$ a non-negative integer. In the limit $|\bar{R}|\gg{m}^2$ -- which holds valid for the whole cosmic history up until the present day in the models to be considered -- the scalar field can be written as
\begin{equation}\label{eq:sc1}
   f_R \simeq -\left| f_{R0} \right|\left(\frac{\bar{R}_0}{R}\right)^{n+1},
\end{equation}
where $\bar{R}_0$, $f_{R0}$ are the $z=0$ values of $\Bar{R}$ and $\bar{f}_R$ respectively. In this work, we set $n=1$ \citep[different values of $n$ produce models that behave qualitatively similarly to our choice of $n$, see, e.g.,][]{Ruan:2021wup}. The chameleon screening mechanism works because Eq.~\eqref{eq:fR} dynamically drives $f_R$ towards $0$ (and hence $\delta R=R-\bar{R}\approx-8\pi G\bar{\rho}_{\textrm{m}}\delta$ which holds for GR) so that the fifth force, being the gradient of $f_R$, becomes vanishingly weak in regions of high density. The mechanism does not necessarily always work everywhere, and if it fails, then the solution is driven to a regime where $R(f_R)\approx\bar{R}$, according to which Eq.~\eqref{eq:fR} predicts that the fifth force has a strength which is $1/3$ of that of the Newtonian gravity. The latter can happen in the regions of low density which are abundant in the Universe, making such regions a useful place to look at when seeking to constrain this MG model.

The choice of the nDGP and $f(R)$ models as test cases in this work (and many other) is motivated by a few considerations. First, chameleon and Vainshtein screening mechanisms are both interesting ways to hide the fifth force through nonlinear self-interaction of the scalar field and its interactions with matter, making them potentially essential building blocks of other, more realistic, MG models. Second, these models are representative in their physical properties and cosmological behaviours, and thus can be useful toy models to gain useful physical insight. Third, these models have practically identical expansion history as standard $\Lambda$CDM (with the same $\Omega_{\textrm{m}}$), so that when investigating them one could easily single out the effect of the fifth force and its screening. For $f(R)$ gravity, this is because any variation of it that predicts an expansion history significantly different from that of $\Lambda$CDM would automatically lose the screening property \citep[see, e.g.,][]{Wang:2012kj,Ceron-Hurtado:2016jrp} and become incompatible with existing GR constraints; for nDGP it is instead a choice: the model on its own cannot predict accelerated expansion and for this some dark energy component has to be included, so one may as well assume that such a dark energy component leads to a $\Lambda$CDM expansion history for the sake of simplicity. Finally, these models are well studied in the literature, so that new results can easily be compared with previous ones, which is useful especially when one's objective is to test the constraining power of a new observational statistic as compared with existing ones, as we are doing here.

This work is based on the \textsc{forge}-\textsc{bridge} suite of simulations, which consist of 50 $f(R)$ models sampled from the 4-dimensional parameter space spanned by $\Omega_{\rm m}, H_0$, $S_8$ (or equivalently $\sigma_8$, with $S_8\equiv\sigma_8(\Omega_{\rm m}/0.3)^{0.5}$), $f_{R0}$, and 50 nDGP models sampled from the 4-dimensional space spanned by $\Omega_{\rm m}, H_0, \sigma_8$ and $H_0r_{\rm c}$. For the former, we vary $|f_{R0}|$ in the range $\left[10^{-7.0}, 10^{-4.5}\right]$. For the latter, we choose $H_0r_{\rm c}$ values between $0.25$ and $10$ \citep[see, for example, Table A1 of][for the values of the cosmological and MG parameters]{Harnois-Deraps:2022bie}. The parameters of the $f(R)$ and nDGP models can also be found in Fig.~\ref{fig:hypercube}: note that these two suites of simulations share the same parameter values of $\Omega_{\rm m}, H_0$ and $S_8$. The MG parameter ranges sampled here correspond to models that are slightly stronger than those ruled out by current observations \citep[e.g.][]{X.Liu2016}, down to models that are weak enough to be nearly consistent with GR.

In general, for fixed cosmological parameters, e.g., $H_0, \Omega_{\rm m},\sigma_8$, increasing $|f_{R0}|$ and decreasing $H_0r_{\rm c}$ would increase the deviation from GR for $f(R)$ gravity and nDGP, respectively. In the former case, it is because $f_{R0}$ is the background value of the scalar field today, and sets the boundary condition of the field far away from astronomical objects; a larger $|f_{R0}|$ makes it harder for the field to be dynamically driven to $\approx0$ near these objects, hence making the chameleon screening not as effective. For nDGP, this is because a smaller $r_{\rm c}$ makes $\beta(a)$ smaller [cf.~Eq.~\eqref{eq:beta_dgp}] and $\boldsymbol{\nabla}^2\varphi$ larger in amplitude; if $\beta$ is close to unity, e.g., as in the late Universe, it also makes the nonlinear terms in Eq.~\eqref{eq:phi_dgp} less significant, weakening the Vainshtein screening.

\section{Simulations and Mock Lensing Maps}
\label{sec:numerics} 

Cosmological simulations are a crucial tool for making accurate predictions of the large-scale structure of the Universe down to the small, nonlinear, scales where perturbation theory fails. Due to the inherently nonlinear nature of the screening mechanisms in the MG models considered here, the case for employing simulations in their studies is even more compelling. Simulation codes for MG models have developed rapidly in the last decade or so \citep[e.g.,][which is inevitably a list far from complete]{Oyaizu:2008,Li:2009sy,Schmidt:2009,Chan:2009ew,Zhao:2011,Li:2012,Li:2013,Puchwein:2013,Llinares:2014,Arnold:2019,Hernandez-Aguayo:2020kgq,Ruan:2021wup,MG-GLAM2}.

Due to their high cost, until very recently simulations for MG have only been done on a relatively small selection of special cases, usually with cosmological parameters fixed to their best-fit values. While useful for gaining some insight into the qualitative behaviours of the models, this approach is limited in that it fails to capture the cosmology dependence of the predicted observables. This issue has been tackled in recent years by efforts to run simulation suites that cover a range of cosmological parameters and construct emulators \citep[e.g.,][]{Heitmann:2006,Heitmann:2009,Heitmann:2010,Heitmann:2014} for the observables \citep[see, e.g.,][for some recent examples of this line of work in the context of modified gravity models]{Winther:2019,Ramachandra:2020,Arnold:2022,Harnois-Deraps:2022bie,Saez-Casares:2023olw,Mauland:2023pjt,Ruan:2023mgq}. The emulators for various observables and physical quantities constructed this way are accurate, allowing one to make predictions for arbitrary values of the cosmological parameters.

In this work, we will follow the same emulator approach, but construct and apply emulators for different observables from those studied previously. Our numerical pipeline is as follows. First, we use the \textsc{forge} and \textsc{bridge} simulations \citep{Arnold:2022,Harnois-Deraps:2022bie}, which sample the 4 parameters, $[\Omega_{\rm m}, S_8, h]$,and the MG parameter $f_{R0}$ (for $f(R)$ gravity) and $H_0r_{\rm c}$ (for nDGP, using 50 nodes in a Latin hypercube, cf.~Fig.~\ref{fig:hypercube}. A Latin hypercube is a sparse sampling method that allows for an efficient sampling of a higher-dimensional parameter space with minimal sizes of samples (hence minimal required simulation effort), while still allowing for good interpolation accuracy. 
We note that while $\log_{\rm{10}}(f_{R0})$ was sampled using a latin hypercube, we sampled $\log_{\rm{10}}(H_0r_c)$ with values of $H_0r_c$ that give similar matter power spectra to the corresponding $\log_{\rm{10}}(f_{R0})$, that is that the weakest (strongest) nDGP model has roughly the same power spectra enhancement as the weakest (strongest) $f(R)$ model on scales $k>1 h\rm{Mpc^{-1}}$. This was done to allow for more fair comparisons when discussing the two models.
Then, we use the WL maps generated from these simulations, MGLenS \citep{Harnois-Deraps:2022bie}, constructed by projecting particles onto mass shells in past light-cones that are then integrated over along the line of sight to make WL convergence maps. The MGLenS maps assume a source redshift distribution $n(z)$ that is shown in Fig.~\ref{fig:n(z)}, which correspond to five tomographic bins representative of a Stage-IV galaxy survey. Finally, WL statistics are extracted from these maps at each node in the \textsc{forge} and \textsc{bridge} suites, and these are used to train a Gaussian Process Regression emulator. Some further details are given below.

\subsection{The \textsc{forge} and \textsc{bridge} simulations}
\label{sec:mg suites}

For the used simulations, the \textsc{arepo} Poisson solver \citep{Springel:2010mn,Weinberger:2020apjs}, which is used to obtain the standard Newtonian force, is augmented by a multigrid relaxation solver for Eq.~\eqref{eq:fR} in $f(R)$ gravity \citep{Arnold:2019} using the method of \cite{Bose:2016wms} and Eq.~\eqref{eq:phi_dgp} for the nDGP model \citep{Hernandez-Aguayo:2020kgq}.
 
For each of the 50 nodes of the \textsc{forge} and \textsc{bridge} simulations, a pair of runs were carried out, their initial conditions being chosen to minimise the effect of sample variance on the matter power spectrum \citep{Arnold:2022}. Two box sizes were adopted, $500h^{-1}$Mpc and $1.5h^{-1}$Gpc, both with $1024^3$ particles. Together, these give a total of 200 dark-matter-only simulations. In this study, we use only the WL maps constructed using the {high-resolution} runs, with a mass resolution of $m_\mathrm{p} \simeq 9.1 \times 10^9h^{-1}M_\odot$\footnote{This figure is for the fiducial $\Lambda$CDM model, or Node 0, while the actual mass resolution for the MG runs varies across the 50 nodes due to different cosmologies.} and a gravitational softening length of $15\, h^{-1}\mathrm{kpc}$. For all simulations, we fixed the index of the primordial power spectrum and the present-day baryonic density parameter and the dark energy equation of state to $n_\mathrm{s} = 0.9652$, $\Omega_{\rm b} = 0.049199$ and $w_0 = -1.0$.

All simulations start at $z_{\rm ini}=127$, with initial conditions generated using the 2\textsc{lptic} \citep[][]{Crocce:2006mn} code, an initial condition generator based on \textsc{n-genic} \citep[][]{Springel:2005nat} that uses second-order Lagrangian perturbation theory to calculate more accurately  the initial particle displacements given a linear matter power spectrum. The latter, for all models, is generated by the public Boltzmann code \textsc{camb} \citep[][]{Lewis:1999bs}, with the cosmological parameters given in Table A1 of \cite{Harnois-Deraps:2022bie}. Note that we did not modify \textsc{camb} to include any MG effect on the initial power spectrum, because for all the $f(R)$ and nDGP models considered here, the linear power spectra are, to a very good approximation, identical to their $\Lambda$CDM+GR counterparts at the initial time ($z=127$). In other words, all the simulated models, whether GR or MG, share the same primordial amplitude $A_{\rm s}$. We note that the $\sigma_8$ (and $S_8$) values reported in this work correspond to the values for a GR model with given $\Omega_{\rm m}, H_0$ and the chosen $A_{\rm s}$, and they are not the values one would expect from the corresponding MG models. When turning on MG, the excess growth of structures caused by the fifth force generally increases the $\sigma_8$ values, but these values are less accurate reflections of the initial condition (which is also well constrained by CMB), hence our decision to label the simulations by their GR quantities\footnote{We use $\sigma_8$ and $S_8$ instead of $\sigma_8^{\rm GR}$ and $S_8^{\rm GR}$ for notational clarity.}.

We use the publicly-available \textsc{slics} simulations\footnote{\textsc{slics}: slics.roe.ac.uk} to make WL maps needed for estimating covariance matrices. \textsc{slics} is a suite of 954 fully-independent $N$-body runs evolving 1536$^3$ particles in a cubic box of 505 $h^{-1}$Mpc. These simulations all have a fixed cosmological model\footnote{This is a $\Lambda$CDM cosmology with $\Omega_{\rm m} = 0.2905$, $\sigma_8 = 0.826$, $h=0.6898$ and $n_{\rm s} = 0.969$.}, differing only in ICs, offering an ideal tool for estimating sample covariance \citep[see][for further details]{SLICS_1}. We note that, as this assumes $\Lambda$CDM+GR, it could result in under-estimated error bars. A more accurate posterior can be obtained if we estimate the covariance matrix at the best-fit cosmology, GR or MG, but such considerations will be left for future works.

\subsection{MGLens}
\label{sec:mglens}

\textsc{The MGLenS} weak lensing maps are generated using the multiple-plane technique \citep[as the SLICS, described in][]{SLICS_1} by ray-tracing photons in the past light-cone up to redshift $z=3$, with the Born approximation. The mass shells are obtained by projecting the particle data along one of the three Cartesian axes, for a comoving thickness equal to exactly half the simulation box size (i.e., 250 $h^{-1}$Mpc), thereby producing 6 mass planes per particle snapshot. Between 15 and 23 shells are needed to continuously fill the light-cones up to $z=3$, depending on cosmology, selecting at random one of the 6 aforementioned planes and adding a random origin offset to break cross-redshift correlations. Convergence maps are then generated for the five tomographic redshift bins shown in Fig.~\ref{fig:n(z)}, each with $7745^2$ pixels and a sky area of 10 $\rm{deg}^2$. Using different planes and random shifts, 50 pseudo-independent light-cones are constructed for each cosmology and MG model, resulting in a set of 50$\times$50$\times$5$\times$2 = 25,000 maps for analysis \citep[see][for further details]{Harnois-Deraps:2022bie}. We note that these WL maps do not yet account for baryons or galaxy intrinsic alignments. In the future it will be crucial to model these additional components when applying the methodology presented here to real data (which we leave to future studies) in order to achieve unbiased constraints.

\begin{figure}
    \centering
    \includegraphics[width=\columnwidth]{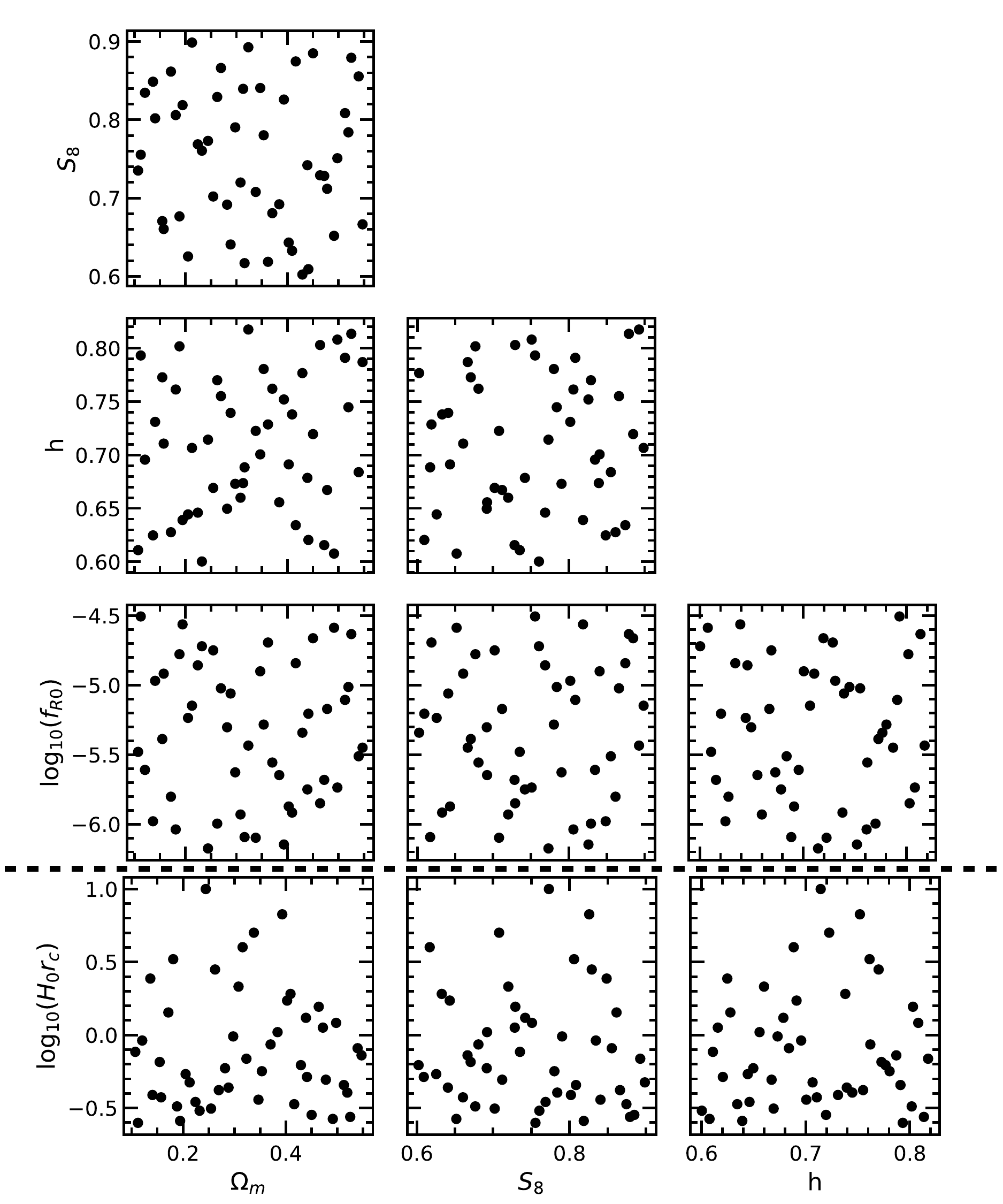}
    \caption{The nodes of the four-dimensional parameter space sampled by two independent simulation suites, \textsc{forge} and \textsc{bridge} \citep{Arnold:2022,Harnois-Deraps:2022bie}. For \textsc{forge}, the $f(R)$ gravity parameter $\log_{10}(|f_{R0}|)$ is sampled (third row), and for \textsc{bridge} the nDGP gravity parameter $\log_{10}(H_0r_{\rm c})$ is sampled (fourth row). The cosmological parameters $\Omega_{\rm{m}}$, $h$ and $S_8$, are sampled identically for the two simulation suites, as indicated by the dashed black line. 
    }
    \label{fig:hypercube}
\end{figure}

\begin{figure}
    \centering
    \includegraphics[width=\columnwidth]{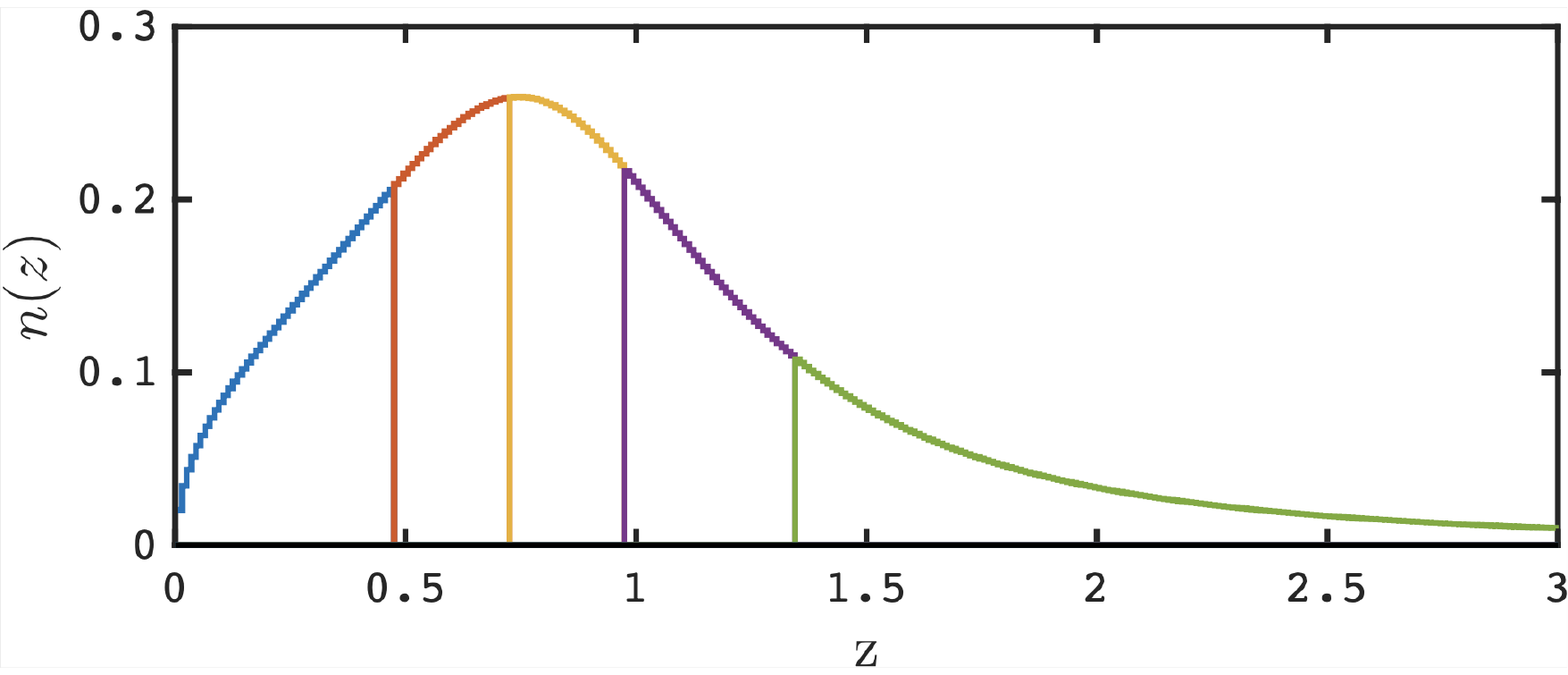}
    \caption{The normalised redshift distribution of source galaxies used to generate five tomographic bins, as indicated by the different colours. }
    \label{fig:n(z)}
\end{figure}

\subsection{Bayesian inference} \label{sec:methodology}

In this work, we use predictions based on the \textsc{forge} and \textsc{bridge} simulations to generate training data, in the form of WL peak statistics, which are used to train a Gaussian Process Regression (GPR) emulator. The trained emulator is then used to predict the peak statistics $\pmb{d}$ for arbitrary parameter values $\pmb{p}$ within the region sampled by the simulation suites, as shown in Fig.~\ref{fig:hypercube}.
This allows us to sample the likelihood of data taken from the \textsc{forge} and \textsc{bridge} simulations and infer the posterior distribution of the cosmological and gravitational parameters.

We use a multivariate Gaussian likelihood, expressed as:
\begin{equation}
    \log(P(\pmb{d}|\pmb{p})) = -\frac{1}{2} \left[\pmb{d} - \mu(\pmb{p})\right]C^{-1}\left[\pmb{d} - \mu(\pmb{p})\right] \, ,
    \label{eq:log likelihood}
\end{equation}
which \JHD{is} sampled with flat priors over the following ranges: $\Omega_{\rm m}$: $[0.10, 0.55]$, $S_8$: $[0.60, 0.90]$, $h$: $[0.60, 0.82]$, and $\log(|f_{R0}|)$: $[-6.17,-4.51]$ for \textsc{forge} and $\log(H_0r_{\rm c})$: $[-0.60,1.18]$ for \textsc{bridge}. In the above expression, $\mu(\pmb{p})$ is the prediction generated by the emulator for a set of parameters $\pmb{p}$, and $C^{-1}$ is the inverse of the covariance matrix. Note that Eq.~\eqref{eq:log likelihood} assumes that the covariance matrix does not depend on the cosmological parameters. 
We use the 954 SLICS WL map realisations (which have the same values of cosmological parameters as the fiducial $\Lambda$CDM cosmology) to estimate the covariance matrix, which we area-rescale (by dividing by the ratio of areas) to a Stage-IV survey area of $18,000$ deg$^{2}$. The joint covariance matrix for the peak probes studied in this work is presented in Appendix \ref{app:cov_mat}. We also multiply the inverse covariance matrix by a debiasing factor $\alpha$, which accounts for the bias introduced when inverting a noisy covariance matrix \citep{Anderson2003,Hartlap2007}, and is given by:
\begin{equation}
    \alpha = \frac{N - N_{\rm{bin}} - 2}{N - 1} \, .
\end{equation}
Here, $N$ is the number of independent realisations used to estimate the covariance of the data vector, and $N_{\rm bin}$ is the number of elements in the data vector.

We use the emulator's prediction of a statistic at a given $\pmb{p}$ as the data $\pmb{d}$. The $\pmb{p}$ values will depend on which node is used as mock data, which we state in the relevant figures, labelled as 'Truth'. This choice is for simplicity and presentation purposes, which ensures that the credible intervals are always centred on the true values of the cosmological parameters, hence allowing for easier comparisons between multiple probes.

\section{results}
\label{sect:results}

This section gives the main results of this work, outlined as follows. We present the measurements and emulation of the WL peak statistics from the \textsc{forge} and \textsc{bridge} simulations in Sections \ref{sec:full n(z)} (without tomography) and \ref{sec:tomo} (with tomography). We then combine these measurements with a GPR emulator to predict these statistics for arbitrary values of the MG parameters, which is presented in the latter half of the respective sections. In Sec.~\ref{sec:forecasts} we present the constraining power of these statistics for a Stage-IV survey, varying simultaneously cosmological and MG parameters. We present such forecasts for MG parameters corresponding to both strong and weak deviations from GR. Finally, in Sec.~\ref{sec:GR} we show the capacity with which our pipeline is able to recover GR ($\Lambda\rm{CDM}$) when departures from GR are allowed. We use this to determine the survey area required to rule out the weakest $f(R)$ and nDGP MG models studied here at $2\sigma$ for the WL peak statistics. 

\subsection{Without tomography}
\label{sec:full n(z)}

In order to measure the WL peak statistics from the \textsc{forge} and \textsc{bridge} MGLenS WL maps for a given node, we first add GSN to each of the maps, as described by Eq.~\eqref{eq: GSN gaussian}. Then, the maps are smoothed using a Gaussian filter with a standard deviation of $1$ arcminute. From each of the smoothed noisy maps, we identify WL peaks as pixels with convergence values larger than that of their eight neighbours, which correspond to local maxima. Once the peaks are identified, we record their positions and amplitudes to generate a WL peak catalogue. This results in 50 peak catalogues, one for each WL map. From each peak catalogue we compute two WL peak statistics, the peak abundance (PA) and the peak 2PCF. We first discuss the results of the peak statistics as measured from convergence maps generated from the full $n(z)$ distribution shown in Fig.~\ref{fig:n(z)}.

\begin{figure*}
    \centering
    \begin{subfigure}{\columnwidth}
        \includegraphics[width=\columnwidth]{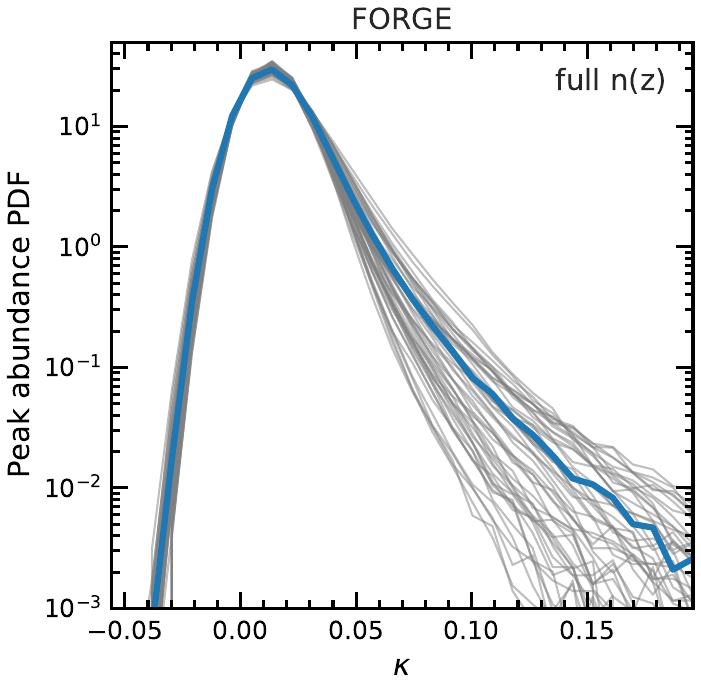}
    \end{subfigure}
    \begin{subfigure}{\columnwidth}
        \includegraphics[width=\columnwidth]{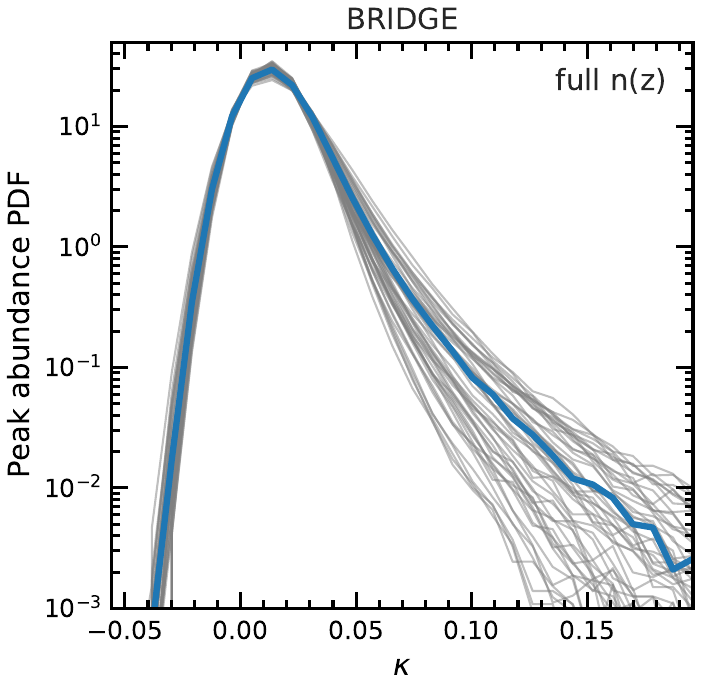}
    \end{subfigure}
    \caption{The weak lensing peak abundances measured from the full n(z) (non-tomographic) weak lensing convergence maps. The two panels correspond to the measurements from the \textsc{forge} (left) and \textsc{bridge} (right) simulation suites. The peak abundance for the $\Lambda \rm{CDM}$+GR node is shown by the blue curves.}
    \label{fig:PA train notomo}
\end{figure*}

For the WL peak abundance we calculate the normalised histograms of the peak amplitudes for each map, giving 50 measurements of the WL PA, from which we compute the mean and standard error. This process is repeated for each node in the \textsc{forge} and \textsc{bridge} simulation suites. The resulting mean PAs for \textsc{forge} and \textsc{bridge} are shown in Fig.~\ref{fig:PA train notomo}. The figure shows that for both the $f(R)$ (left panel) and nDGP (right panel). The peak abundances are negatively skewed in both cases, and the distributions peak at just above the mean convergence ($\kappa=0$) at $\kappa \approx 0.01$. It is clear that it is the skewness that varies more strongly across the different nodes, whereas the peak of the distribution is always near a fixed $\kappa$ value, and varies only slightly in amplitude. It is important to highlight however that the uncertainty on the peak abundance increases dramatically with $\kappa$, which will directly counteract the greater variation with respect to cosmology when testing the cosmological information content of the WL PA. Finally, the figure shows that it is the positive peaks ($\kappa$ > 0), that vary more strongly than the negative peaks ($\kappa$ < 0), as the cosmological parameters change. The fiducial $\Lambda\rm{CDM}$+GR node is shown in blue.

For the peak 2PCF, we impose an additional $\kappa$ threshold on each peak catalogue, with peaks below this threshold removed. We do this for a range of $\kappa$ thresholds of $\kappa =$ $[0.01,0.02,0.03,0.04,0.05]$, resulting in five trimmed peak catalogues for each of the 50 realisations of a given node. Next, the 2PCF for a given trimmed peak catalogue is computed based on Eq.~\eqref{Eq: LS estimator}. We use the $\theta$ range $[0.1,1]$ deg, as smaller scales are limited by the map resolution, and larger scales by are limited by the WL map size. As described in \cite{Davies2019}, we compute the mean 2PCF by taking the average $N_D$, $DD$, and $DR$, and then using these values in conjunction with Eq.~\eqref{Eq: LS estimator}. We use this approach instead of computing $50$ 2PCFs (one for each realisation) and then taking the mean, as this would lead to biased measurements of the peak 2PCF and its corresponding uncertainty \citep{Davies2019}. We therefore calculate the error on the mean using jackknife sampling, where we repeat the previous step $50$ times, but each time we omit a different peak catalogue, to give $50$ measurements of mean $\xi$ values computed from $49$ peak catalogues. The standard error is then computed from the standard deviation of the $50$ jackknifed $\xi$ measurements. 

\begin{figure*}
    \centering
    \begin{subfigure}{\columnwidth}
        \includegraphics[width=\columnwidth]{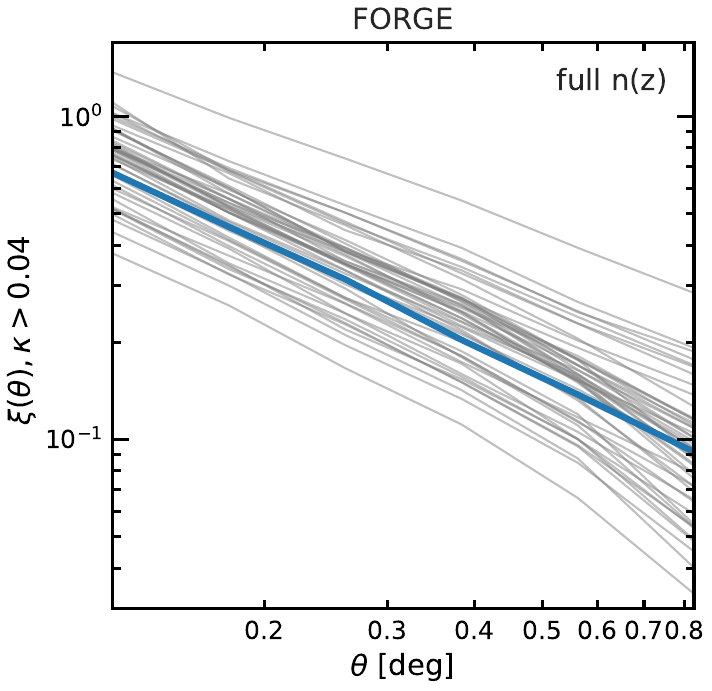}
    \end{subfigure}
    \begin{subfigure}{\columnwidth}
        \includegraphics[width=\columnwidth]{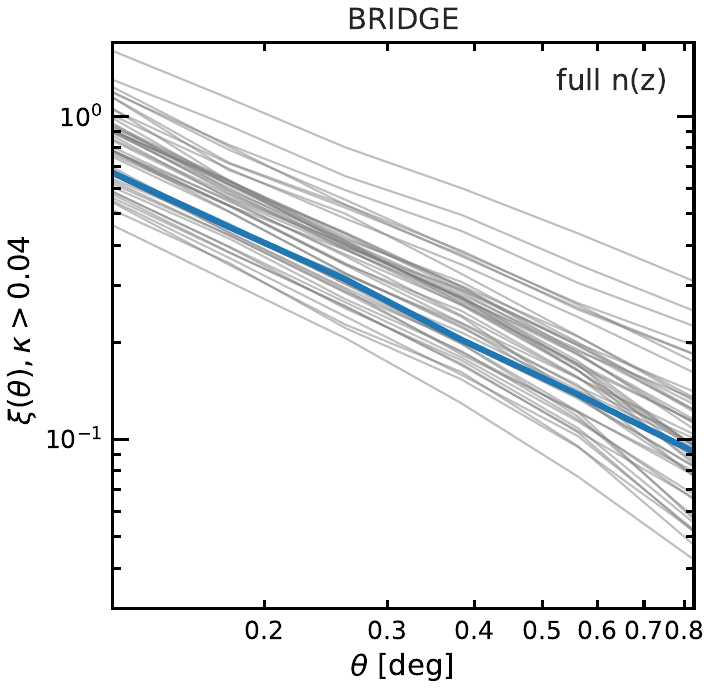}
    \end{subfigure}
    \caption{Same as Fig.~\ref{fig:PA train notomo} except for the peak 2PCF for $\kappa > 0.04$. }
    \label{fig:P2PCF train notomo}
\end{figure*}

The peak 2PCF measurements for the full $n(z)$ are shown in Fig.~\ref{fig:P2PCF train notomo} for peak catalogues with $\kappa > 0.04$. We show only the results for a single $\kappa$ threshold here to illustrate the variance of the peak 2PCF across the \textsc{forge} and \textsc{bridge} nodes. For detailed studies on the behaviour of the peak 2PCF when varying the peak threshold, we refer the readers to \cite{Davies2019,Davies2020b}, as the results from \textsc{forge} and \textsc{bridge} follow qualitatively the same behaviour. Increasing the $\kappa$ threshold increases the amplitude of the peak 2PCF, with negligible evolution of the slope: we see that the 2PCFs follow a power-law, and that their amplitude varies by roughly a factor of two, across the range of parameters sampled by \textsc{forge} and \textsc{bridge}. Again, the blue curve indicates the measurement for the fiducial $\Lambda\rm{CDM}$+GR case.

Next we use the statistics presented in Figs.~\ref{fig:PA train notomo} and \ref{fig:P2PCF train notomo} as training data for the GPR emulator, following the same method from \cite{davies/etal:2021}. The accuracy of the emulator used in this analysis is presented and discussed in Appendix \ref{app:emm_acc}.
Before running a full Stage-IV forecast, we first use the emulator to investigate how these statistics depend on the MG parameters. This offers useful insight into the physics underpinning these statistics, which is presently not possible to accurately calculate with a fully theoretical model, as these statistics are a consequence of highly nonlinear features of the large-scale structure.

\begin{figure*}
    \centering
    \begin{subfigure}{\columnwidth}
        \includegraphics[width=\columnwidth]{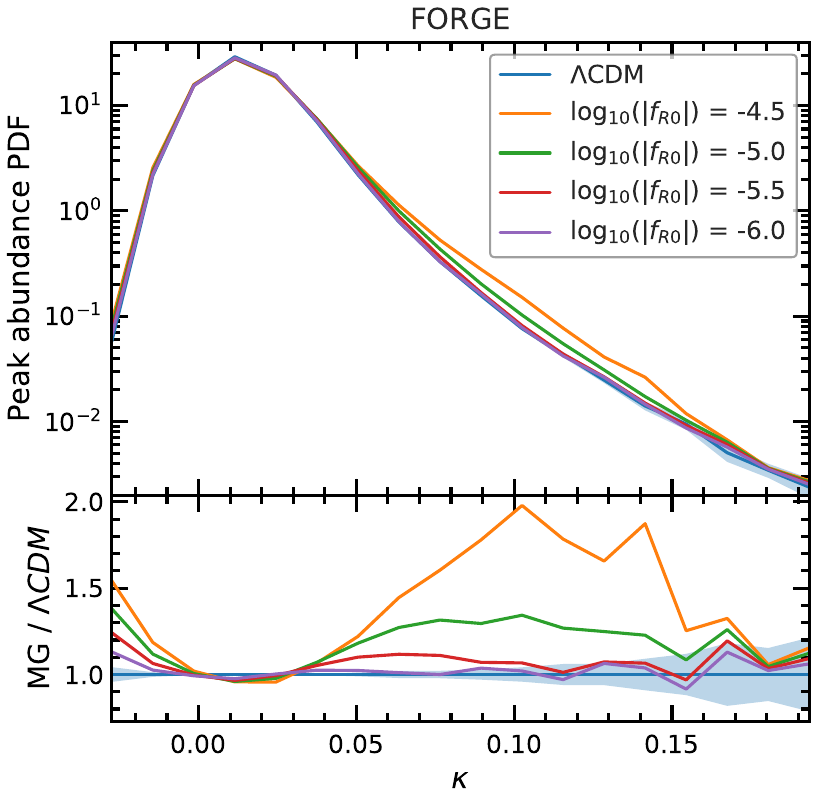}
    \end{subfigure}
    \begin{subfigure}{\columnwidth}
        \includegraphics[width=\columnwidth]{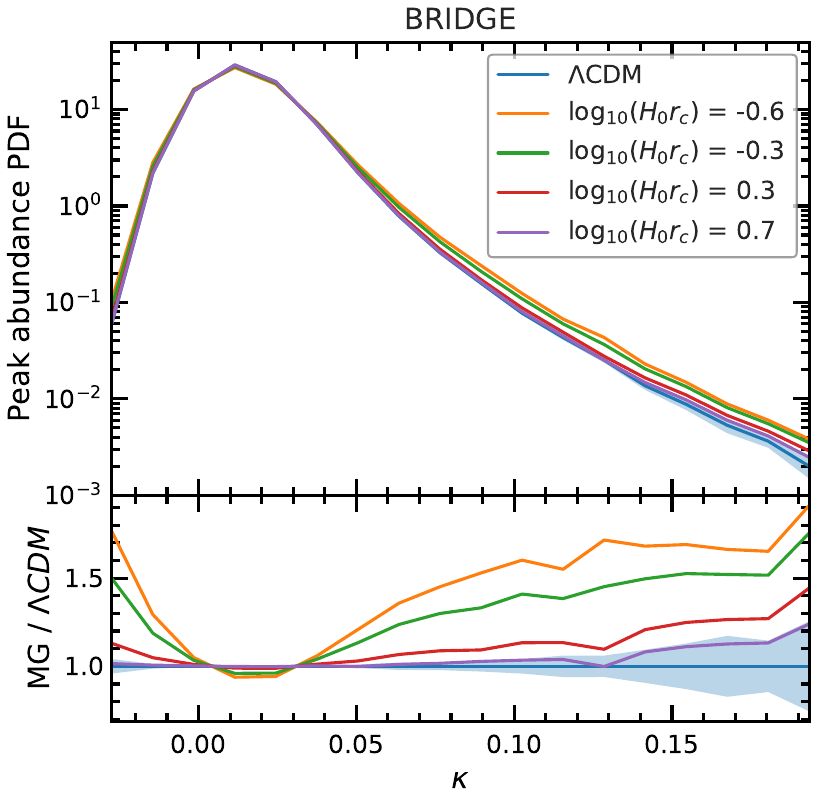}
    \end{subfigure}
    \caption{Upper panels: emulation of the non-tomographic WL peak abundance for a range of modified gravity parameters. The left panels corresponds to the \textsc{forge} emulation, where $\log(|f_{R0}|)$ is varied, and the right panels corresponds to the \textsc{bridge} emulation where $\log(H_0r_{\rm c})$ is varied. The values of the MG parameters are indicated by the panel legends. The blue curves represent the emulated $\Lambda \rm{CDM}$ GR non-tomographic WL peak abundances ($\log(|f_{R0}|) = -\infty$ for $f(R)$ and $\log(H_0r_{\rm c}) = \infty$ for nDGP). Lower panels: the ratio of the emulated peak abundances relative to the $\Lambda \rm{CDM}$ GR case. The shaded blue region indicates the $1-\sigma$ standard errors.
    }
    \label{fig:PA emu notomo}
\end{figure*}

Fig.~\ref{fig:PA emu notomo} shows the emulated peak abundances generated from the \textsc{forge} (left) and \textsc{bridge} (right) emulators. The cosmological parameters are fixed to the fiducial values of the $\Lambda$CDM+GR node (see \cite{Harnois-Deraps:2022bie}), and only the MG parameters are varied, as shown by the legends in the upper right corner of each panel. The ratio of the peak abundances relative to the prediction for the $\Lambda\rm{CDM}$+GR node by the same emulator are displayed in the lower sub panels. The shaded blue region corresponds to the standard error for the $\Lambda\rm{CDM}$ simulated peak abundance. 

The figure shows that both MG models produce an enhancement of the WL peak abundance for $\kappa < 0$. This follows intuitively from the fact that $\kappa < 0$ corresponds to underdense lines-of-sight, where the fifth force in the MG models is unscreened, so that matter in central underdense regions is more strongly evacuated towards neighbouring regions, further lowering the $\kappa$ values in the former.

The enhancement drops off for both models as $\kappa$ increases towards 0, and then becomes a suppression relative $\Lambda\rm{CDM}$+GR, before returning to an enhancement at $\kappa \approx 0.03$. The reason for this suppression results from the fact that the MG models are enhancing structure formation, which results in both more underdense regions and greater over densities. This in turn means that there are fewer structures close to the mean value of the density field, which here corresponds to $\kappa = 0$, and hence the suppression of the peak abundance in this regime is essentially a consequence of the conservation of matter. 

In addition, both models display an increase in the WL peak abundance for $\kappa > 0.03$. Again, this is due to the enhanced structure formation in these models caused by the fifth force. However, the behaviour here is slightly more complex compared to the $\kappa < 0$ case, as this regime corresponds to overdense lines-of-sight, and along some of these the fifth force is expected to be screened. WL peaks are produced by a combination of chance alignments of haloes along the line-of-sight, massive dark matter structures, and their coupling to the large-scale modes of the matter field \citep[e.g.,][]{Yang2011,J.Liu2016, Wei2018, Sabyr2022}. This means that enhancement in the WL peak abundance for $\kappa > 0$ is tightly linked to how these models alter the halo mass function \citep[e.g.,][]{Vogt2024}, and the matter power spectrum \citep{Arnold:2022}. Therefore, the observed enhancement is produced by the following factors. First, there is an increased chance of coincidental alignments of low-mass haloes along the line of sight, due to the small haloes being more numerous in MG relative to $\Lambda\rm{CDM}$+GR. Second, for peaks (dominantly) produced by individual massive haloes, the corresponding $\kappa$ values are enhanced, due to the massive halo also having a greater mass in the corresponding MG model. The increase in mass is produced by increased accretion of matter onto the halo, which is driven by the fifth force.

The two sub-panels also illustrate how the two MG models alter the peak abundance in different ways. 
This is evident from the fact that for nDGP, there is a continuous enhancement of WL peaks for large $\kappa$ values, whereas the enhancement drops off as $\kappa$ increases for $f(R)$ gravity. This is because for $f(R)$ gravity, the enhanced growth of very massive haloes is suppressed by chameleon screening, which causes a relatively short range of the fifth force around such haloes so that the halo mass function goes back to the GR prediction for them; for nDGP, in contrast, the fifth force is screened only on small scales, while it has full strength on large scales, so that more matter can be accreted onto very massive haloes from the rich reservoir of matter around them \citep[see, e.g., Fig.~9 and Fig.~11 respectively of][for the behaviour of the halo mass function enhancment in these two models]{Ruan:2021wup,MG-GLAM2}.
This shows that the WL peak abundance is not only sensitive to the presence of MG, but also to the type of screening mechanism. 

\begin{figure*}
    \centering
    \begin{subfigure}{2\columnwidth}
        \includegraphics[width=\columnwidth]{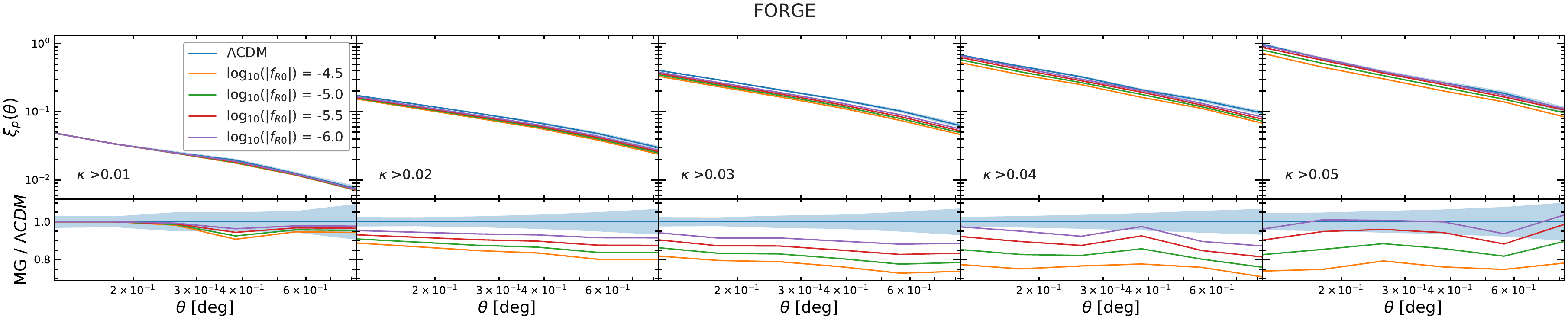}
    \end{subfigure}
    \begin{subfigure}{2\columnwidth}
        \includegraphics[width=\columnwidth]{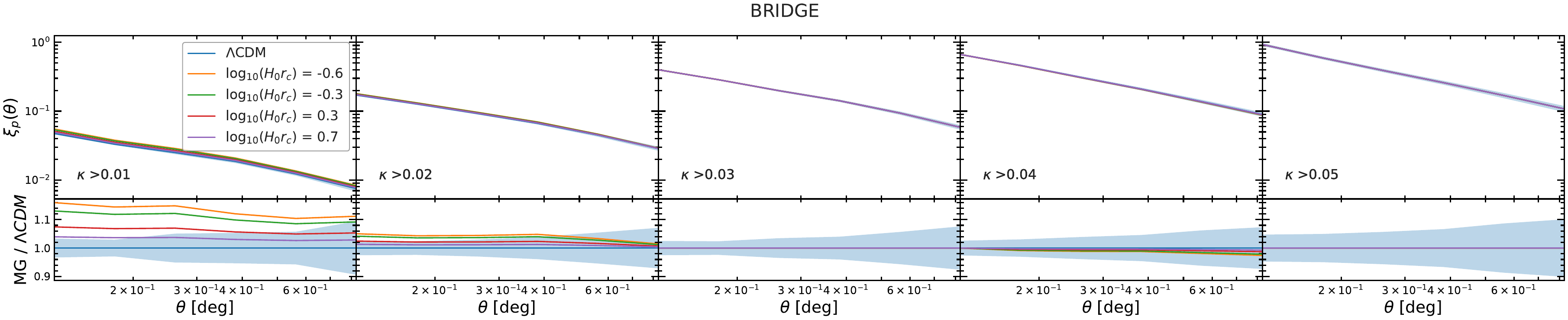}
    \end{subfigure}
    \caption{Emulation of the non-tomographic WL peak 2PCFs. The upper and lower rows correspond to $f(R)$ and nDGP respectively, where the strength of the MG parameter is varied, as shown by the legend in the leftmost sub-panels. Each column shows the WL peak 2PCF for a given $\kappa$ threshold, shown by the labels in the lower left of the upper sub-panels. The lower sub-panels show the ratio of the WL peak 2PCF for a given MG parameter relative to the $\Lambda\rm{CDM}$ case. The shaded blue region indicates the $1-\sigma$ standard errors for $\Lambda\rm{CDM}$.   }
    \label{fig:P2PCF emu notomo}
\end{figure*}

Fig.~\ref{fig:P2PCF emu notomo} shows the same method from \ref{fig:PA emu notomo} but repeated for the emulated WL peak 2PCFs, without using tomography. The columns correspond to each of the five $\kappa$ thresholds, as shown in the legends. The top and bottom rows show the \textsc{forge} and \textsc{bridge} emulator predictions for the $f(R)$ and nDGP peak 2PCFs respectively.
The figure shows that the amplitude of the clustering signal increases with the $\kappa$ threshold, as reported in previous works \cite{Davies2019,davies/etal:2021}, indicating a qualitatively similar behaviour in MG models as in $\Lambda\rm{CDM}$. For $f(R)$ gravity, the lower sub-panels show that the clustering on WL peaks is suppressed relative to $\Lambda\rm{CDM}$, where this suppression increases as the deviation from GR increases. Furthermore, for models which deviate more strongly from GR, such as $\log(|f_{R0}|) = -4.5$, the amount by which the clustering is suppressed relative to $\Lambda\rm{CDM}$ also increases as the $\kappa$ threshold increases, whereas for weaker deviations from GR, the suppression stops growing and begins to decrease again past $\kappa > 0.03$. This exemplifies the fact that there is complementary information between the different peak 2PCFs in MG, as the response to varying the MG parameter changes in each case, both quantitatively and qualitatively. This behaviour has also been studied in the case of $w\rm{CDM}$ \citep{davies/etal:2021}, however, as we will show later, the information gained from this complementarity is less dramatic for MG, which is largely driven by strong degeneracies between the MG and cosmological parameters.

For the case where the nDGP MG parameter is varied (bottom row), we find a distinctly different response in the peak 2PCF, where decreasing the value of $H_0r_{\rm c}$ leads to an enhancement in peak clustering relative to $\Lambda\rm{CDM}$+GR. However, this behaviour is only present for low $\kappa$ thresholds, where for $\kappa > 0.03$ and above, the emulator does not show a dependence of the peak 2PCF on $H_0r_{\rm c}$ at a statistically significant level. 

The different behaviour of the peak 2PCF in the $f(R)$ and nDGP models is again rooted in their different screening mechanisms. In the $f(R)$ case, where the chameleon mechanism screens the fifth force, small, less clustered haloes have increased mass (and hence higher $\kappa$) relative to their $\Lambda\rm{CDM}$+GR counterparts. When the $\kappa$ threshold is applied to the peak catalogues, more haloes with a lower clustering signal remain in the catalogue compared to $\Lambda\rm{CDM}$, resulting in a suppression of the 2PCF. In the case of the Vainshtein mechanism in the nDGP model, the fifth force is able to operate on larger scales than in the $f(R)$ case. This leads to an enhancement of structure formation in larger-scale modes relative to $f(R)$, which further boosts the clustering of already clustered objects, which results in a 2PCF with increased amplitude. However, the scale at which the fifth force is unscreened here is not large enough to alter structure formation in the most massive haloes, which is why the enhancement fades away as the $\kappa$ threshold increases. This can also be interpreted in terms of the matter power spectrum as shown in \cite{Gupta2023}, where it is clear that nDGP provides a boost in power at larger scales relative to $\Lambda\rm{CDM}$, and then drops off at small scales, compared to $f(R)$ gravity.

\subsection{With Tomography}\label{sec:tomo}

Next, we repeat the analysis from the previous section, except now we study the properties of the WL peak statistics as a function of the source redshift tomographic bin. Exploiting the tomographic information is a commonly-used approach with weak lensing statistics \citep[e.g.,][]{des/kids:2023}, which provides additional cosmological information to the non-tomographic case. Here, we study how WL peak statistics change as a function of the source redshift, and study the impact of MG relative to $\Lambda\rm{CDM}$. We note that the WL $\kappa$ maps in single tomographic bins are more noisy, as fewer source galaxies contribute to the observed signal, however the increase in noise is more than offset by the gain in cosmological information that can be extracted from the redshift evolution. 

\begin{figure*}
    \centering
    \begin{subfigure}{2\columnwidth}
        \includegraphics[width=\columnwidth]{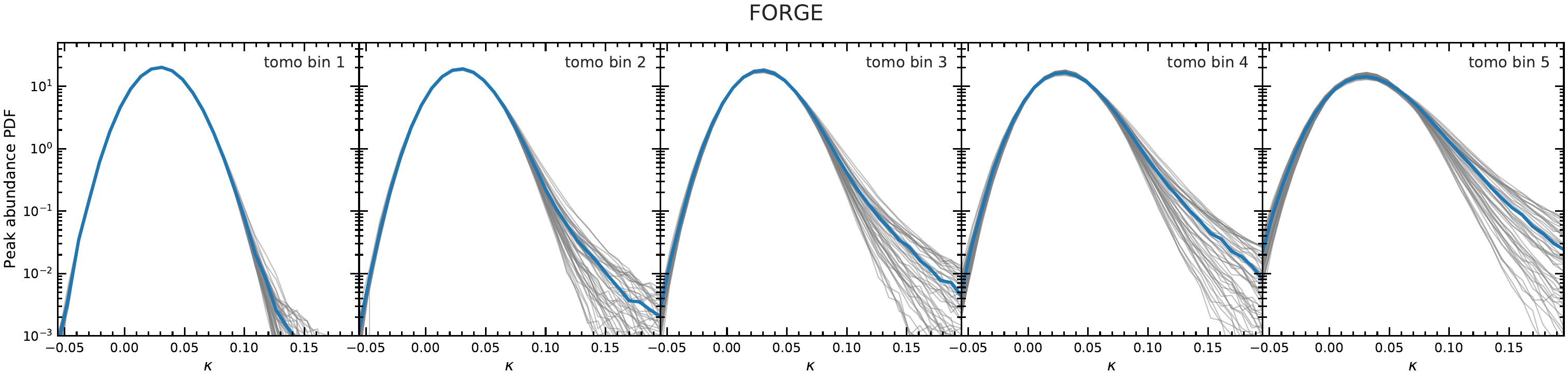}
    \end{subfigure}
    \begin{subfigure}{2\columnwidth}
        \includegraphics[width=\columnwidth]{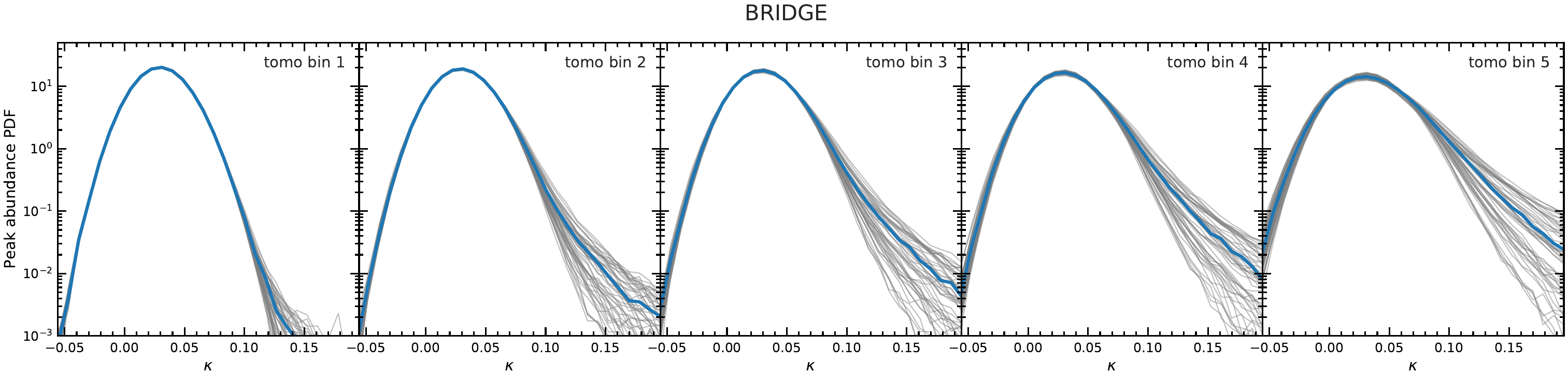}
    \end{subfigure}
    \caption{The same as Fig.~\ref{fig:PA train notomo} but for the tomographic case. The top and bottom rows correspond to the WL peak abundances measured in \textsc{forge} and \textsc{bridge} respectively. Each column corresponds to a given tomographic bin (see Fig.~\ref{fig:n(z)}), as indicated by the label in the top right corner of each panel. }
    \label{fig:PA train tomo}
\end{figure*}

Fig.~\ref{fig:PA train tomo} shows the tomographic WL peak abundances as measured from the \textsc{forge} (top row) and \textsc{bridge} (bottom row) nodes. Each column corresponds to a different tomographic bin, as shown by the labels in the top right of the sub-panels. The scatter in the MG peak abundances (grey lines) appears to be visually very similar for both the \textsc{forge} and \textsc{bridge} simulations. This is the training data for the emulator, which will allow us to explicitly extract the impact of the MG parameters on the WL peak abundance shortly. As the source redshift of the tomographic bin increases, all of the peak abundances become more non-Gaussian, with increased positive skewness. This is due to the following reason. Whilst the GSN level is the same in each tomographic bin, the signal induced by the matter along the line of sight increases with higher source redshift. 

\begin{figure*}
    \centering
    \begin{subfigure}{2\columnwidth}
        \includegraphics[width=\columnwidth]{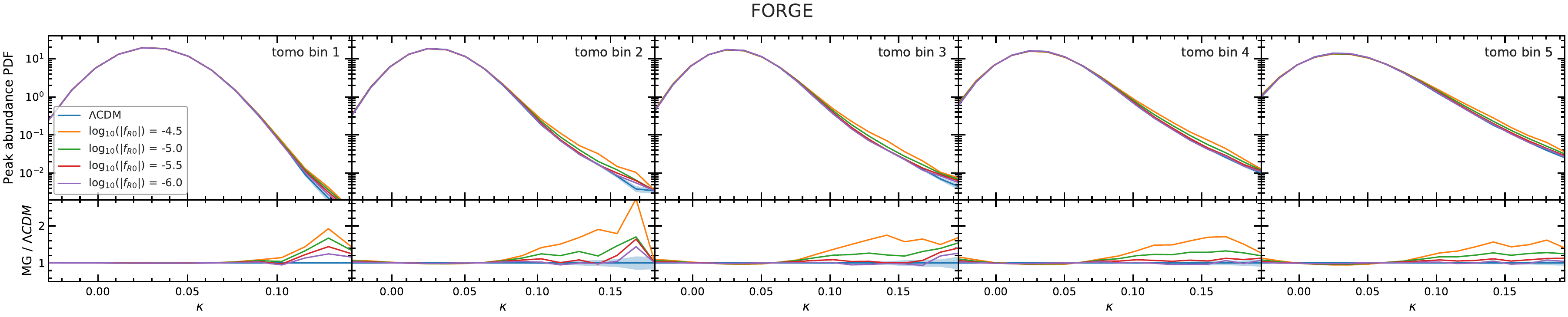}
    \end{subfigure}
    \begin{subfigure}{2\columnwidth}
        \includegraphics[width=\columnwidth]{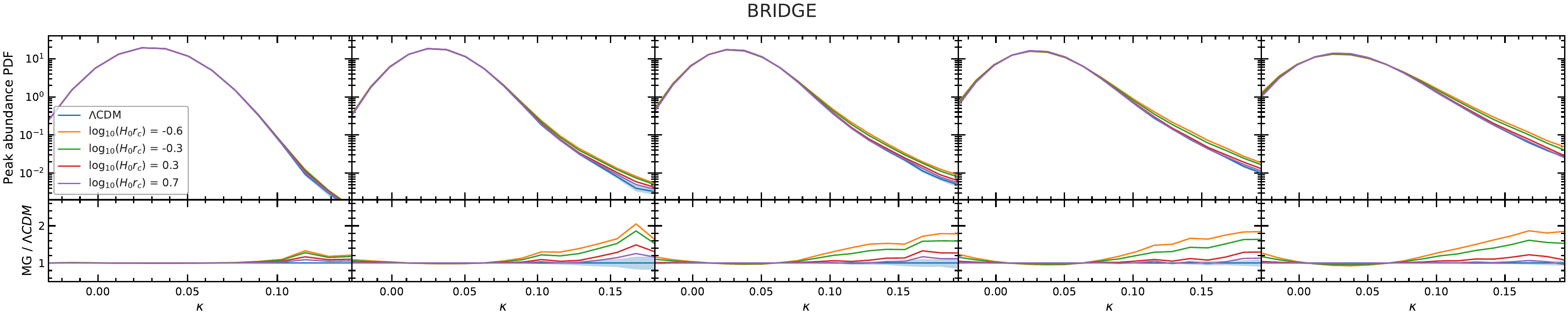}
    \end{subfigure}
    \caption{The same as Fig.~\ref{fig:PA emu notomo} except now the tomographic case is presented. The top and bottom rows are predictions by the WL peak abundance emulators built from the \textsc{forge} and \textsc{bridge} simulations, respectively. Each column corresponds to a given tomographic bin (see Fig.~\ref{fig:n(z)}), as indicated by the label in the top right corner of each panel.}
    \label{fig:PA emu tomo}
\end{figure*}

Fig.~\ref{fig:PA train tomo} shows the emulated tomographic WL peak abundances for \textsc{forge} (top row) and \textsc{bridge} (bottom row), where the columns correspond to different tomographic source redshift bins as in Fig.~\ref{fig:PA train tomo}.

For the $f(R)$ case, the lowest source redshift bin shows the weakest enhancement with respect to the $\Lambda\rm{CDM}$, which falls off at $\kappa = 0.15$ and is poorly sampled due to noise in the training data. As shown by Fig.~\ref{fig:PA train tomo}, peaks in this regime are exceedingly rare.
For the higher source redshift bins, the enhancement induced by $f(R)$ gravity appears to be broadly consistent with eachother, where the second tomographic bin gives the strongest enhancement.
In addition to enhancement at high $\kappa$, there is also an enhancement in the peak abundance for $\kappa < 0$. This follows intuitively from the fact that we expect modifications to structure formation from $f(R)$ to become apparent at low redshift. Since the lensing kernel in Eq.~\eqref{eq:convergence} weights structures at $\chi' = \chi / 2$ the highest, it is natural to find that the highest redshift bins are less sensitive to $f(R)$ than their low redshift counterparts, where MG induced structure formation is stronger. Overall there appears to be a slight trend for a reduction in the peak abundance enhancement induced by $f(R)$ as a function of redshift

Next, for nDGP, we again see that the lowest source redshift bin yields the weakest enhancement from nDGP, due to the high noise contribution here. For the remaining source redshift bins, in contrast to the $f(R)$ case, we see that the enhancement in the peak abundance at high $\kappa$ from nDGP increases monotonically with redshift. This difference is due to the different screening mechanisms in each model, with the Vainshtein screening mechanism allowing the fifth force to be unscreened on larger scales than the chameleon mechanism. For the high source redshift case, since the integral over $\chi$ in Eq.~\eqref{eq:convergence} corresponds to greater distances, more large-scale modes are captured than compared to the low source redshift counterparts, so that the enhancement induced by these modes becomes stronger. Whilst it is true that the same modes are also captured by this redshift bin in the $f(R)$ case, the same behaviour is not observed because the chameleon screening mechanism does not efficiently enhance to power of the large scale modes relative to $\Lambda\rm{CDM}$. The enhancement for $\kappa < 0$ and suppression at $\kappa \approx 0.03$ seen here were explained in Sec. \ref{sec:full n(z)} and observed in  the no-tomographic case.

\begin{figure*}
    \centering
    \begin{subfigure}{2\columnwidth}
        \includegraphics[width=\columnwidth]{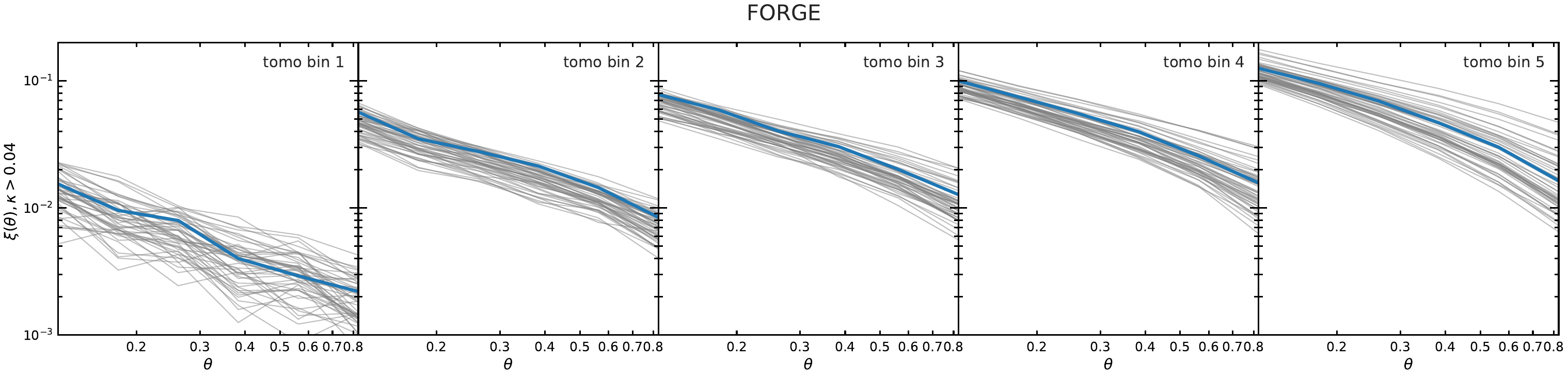}
    \end{subfigure}
    \begin{subfigure}{2\columnwidth}
        \includegraphics[width=\columnwidth]{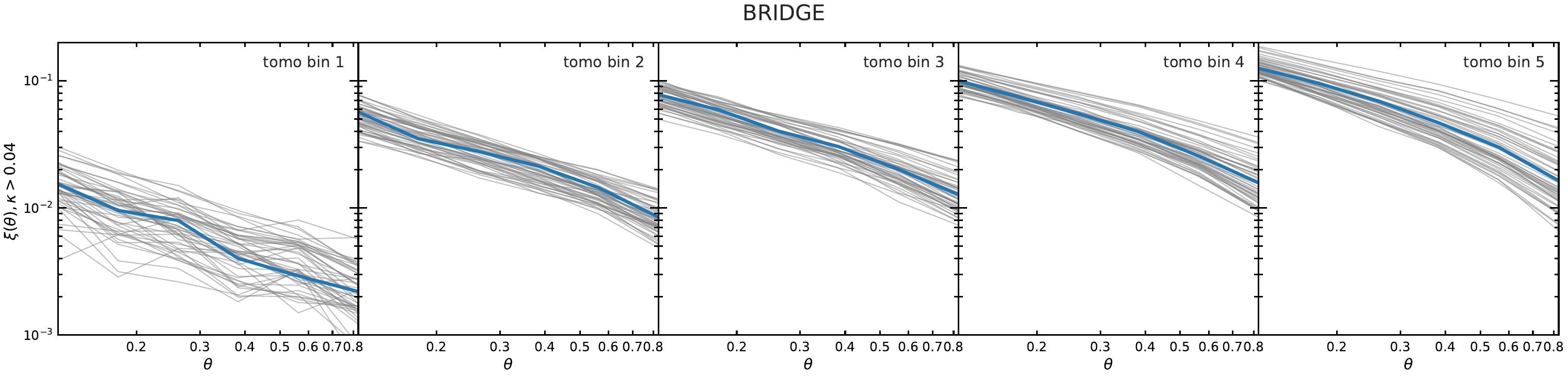}
    \end{subfigure}
    \caption{The same as Fig.~\ref{fig:P2PCF train notomo} except for the tomographic case. The top and bottom rows correspond to the WL P2PCFs measured from \textsc{forge} and \textsc{bridge} respectively. Each column corresponds to a given tomographic bin (see Fig.~\ref{fig:n(z)}), as indicated by the label in the top right corner of each panel. We show on the WL P2PCFs for $\kappa > 0.04$ to exemplify its behavior as a function of the tomographic bin. }
    \label{fig:P2PCF train tomo}
\end{figure*}

Next we consider the WL peak 2PCFs for the tomographic case. Fig.~\ref{fig:P2PCF train tomo} shows the measurements from the \textsc{forge} (top row) and \textsc{bridge} (bottom row) simulations. We have found that the dependence of the WL peak 2PCF on the tomographic bin is qualitatively the same for all $\kappa$ thresholds, and so we only show the 2PCFs for the case with $\kappa > 0.04$ for brevity. The figure shows that the amplitude of the signal increases as the redshift of the sources increases. Additionally, the measurement is noisier for the low redshift tomographic bins. 

Given that the matter auto-correlation function increases in amplitude with time \cite{Matarrese1997}, one might also expect that the WL peak 2PCF exhibits the same behaviour. However, for DM haloes at early times, only the most biased structures have formed, which are highly clustered. As we approach later times, more haloes form, where these haloes are now less biased and hence less clustered. This means that although the clustering of existing haloes is increasing with time, new less clustered haloes are also forming, which dilutes the total clustering signature. In practice, the latter dominates the former, resulting in a decrease in the average halo clustering at later times. This is analogous to the behaviour observed with the WL peak 2PCF, given that there is a strong connection between WL peaks and DM haloes.

It is also important to consider the response of the WL peak 2PCF to GSN. Here, low-redshift bins are more noise dominated than the high-redshift bins. This is due to the integration of matter along a shorter line of sight for low redshift bins, cf.~Eq.~\eqref{eq:conv source} for a more detailed discussion), which results in peaks with a lower signal-to-noise ratio. This means that more spurious peaks are produced. These spurious peaks are not clustered, which leads to a 2PCF with a lower amplitude.

\begin{figure*}
    \centering
    \begin{subfigure}{2\columnwidth}
        \centering
        \includegraphics[width=\columnwidth]{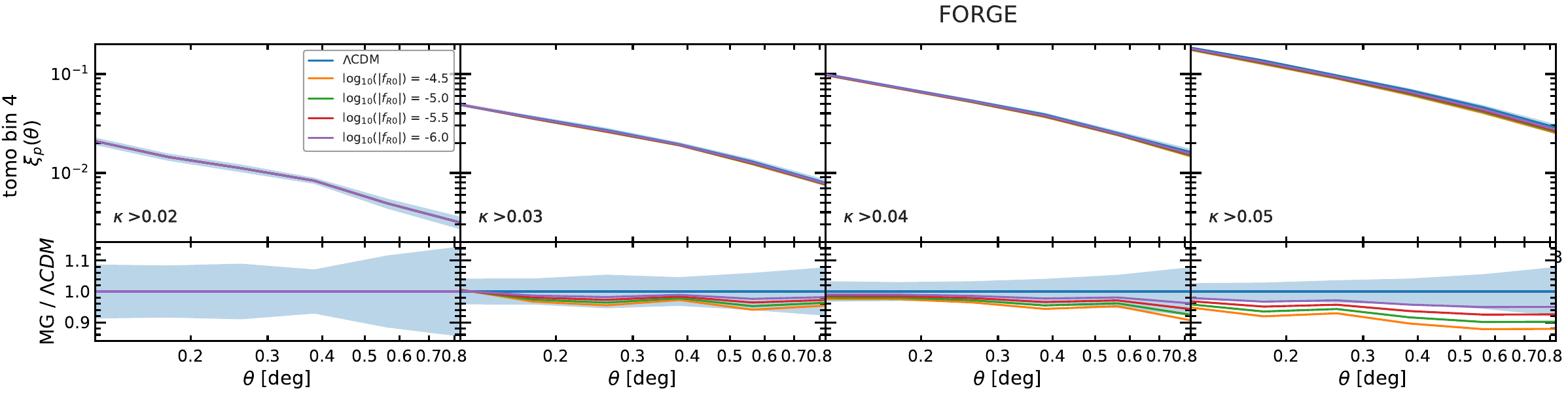}
    \end{subfigure}
    \begin{subfigure}{2\columnwidth}
        \includegraphics[width=\columnwidth]{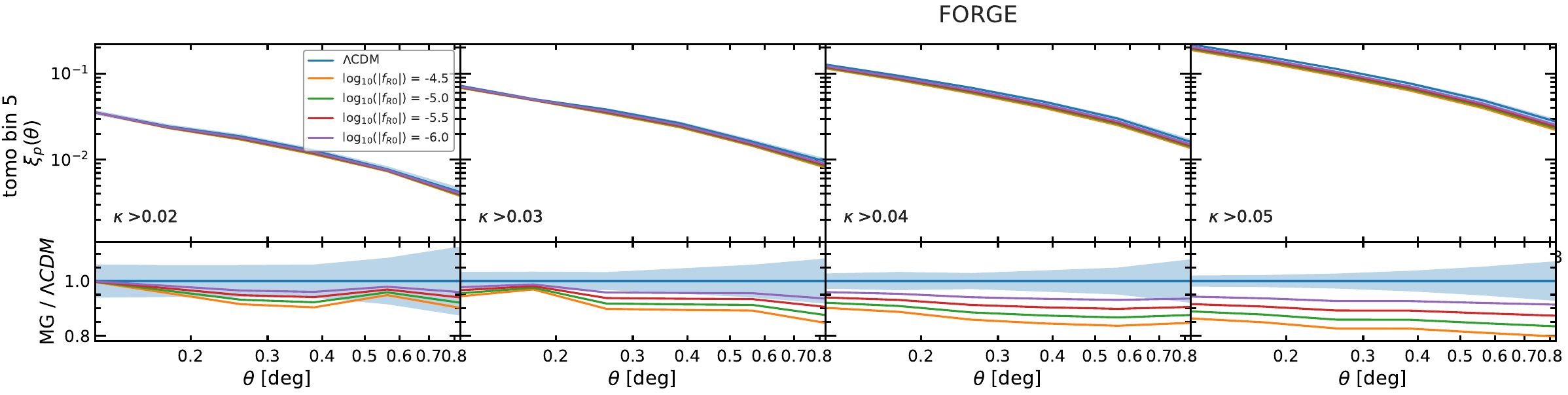}
    \end{subfigure}
    
    \begin{subfigure}{2\columnwidth}
        \centering
        \includegraphics[width=\columnwidth]{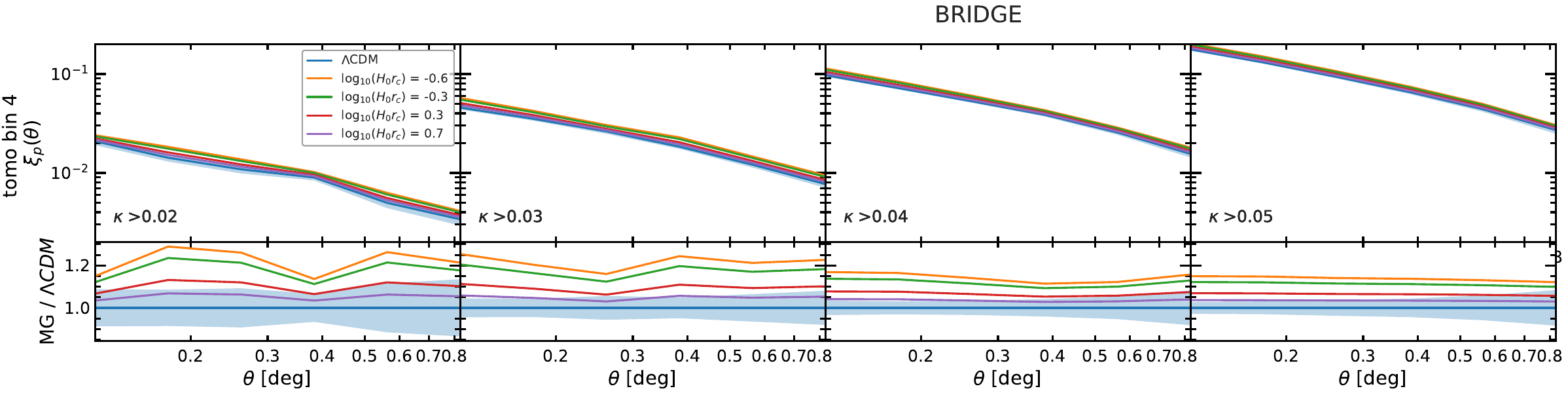}
    \end{subfigure}
    \begin{subfigure}{2\columnwidth}
        \includegraphics[width=\columnwidth]{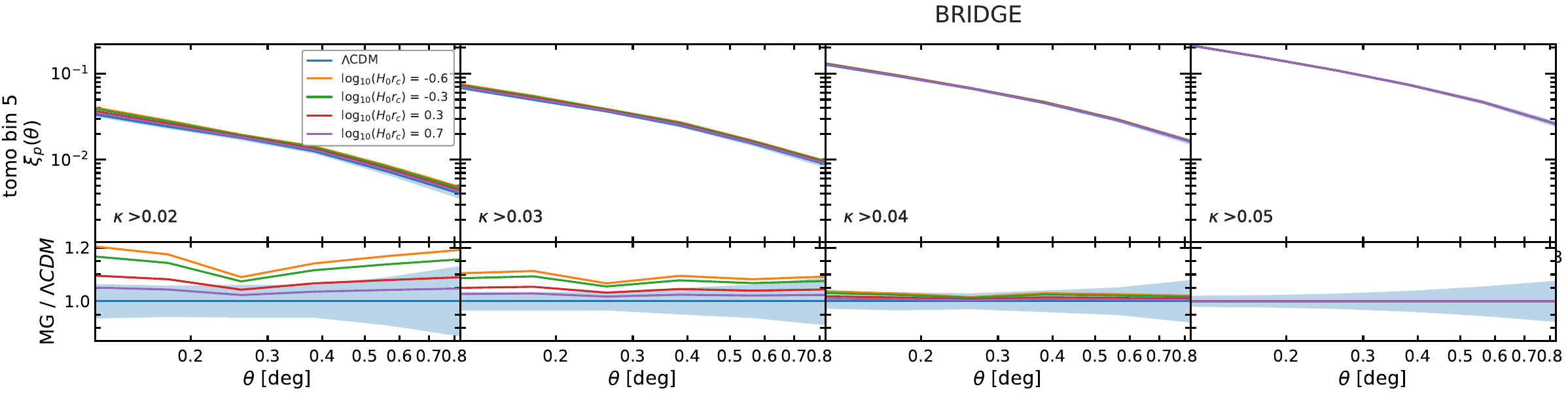}
    \end{subfigure}
    \caption{Emulator predictions of the WL peak 2PCF from \textsc{forge} (top two rows) and \textsc{bridge} (bottom two rows). Results are shown for tomographic bins 4 (first rows) and 5 (second rows). Each column corresponds to a different $\kappa$ threshold, as shown by the labels in the bottom left corner of the upper sub-panels. The lower sub-panels show the ratio of a given WL peak 2PCF relative to the emulator at $\Lambda\rm{CDM}$+GR, where the shaded blue region indicates the associated $1$-$\sigma$ standard error. }
    \label{fig:P2PCF emu tomo}
\end{figure*}

Next we discuss the emulated tomographic WL peak 2PCF and its dependence on MG parameters.
The top two rows of Fig.~\ref{fig:P2PCF emu tomo} show the emulator predictions for tomographic bins 4 (first row) and 5 (second row). 
Note that we haven't included the result for the $\kappa>0.01$ threshold, because the low physical signal and high noise contribution makes it difficult to reliably train the GPR emulator. In addition to this, the impact of MG on the tomographic peak 2PCF is small for the low $\kappa$ thresholds for the non-tomographic case, and as such we do not include the $\kappa > 0.01$ tomographic results. Similar to the non-tomographic case, we can see that the 2PCF is suppressed relative to $\Lambda\rm{CDM}$ as the deviation from GR increases in the $f(R)$ model. The overall magnitude of this suppression is smaller in the tomographic case, which is due to the fact that GSN is more dominant here. As shown by the bottom row, this suppression is greater for the higher redshift tomographic bin, and the corresponding errors are also smaller, where GSN is less dominant.

Next, the bottom two rows of Fig.~\ref{fig:P2PCF emu tomo} show the same results as the top two rows, except for the nDGP model. 
Similar to the non-tomographic case, nDGP gravity produces greater enhancements of the 2PCF for low $\kappa$ thresholds, and this enhancement decreases as the $\kappa$ thresholds increases. The overall magnitude of the 2PCF enhancement is lower for the higher redshift tomographic bin. This is consistent with the fact that the fifth force is more dominant at low redshift.

From Fig.~\ref{fig:P2PCF emu tomo}, it is clear that the $f(R)$ and nDGP models modify the WL peak 2PCF in distinctly different ways: $f(R)$ gravity gives a suppression relative to $\Lambda\rm{CDM}$ that becomes stronger with increasing $\kappa$ threshold and source redshift of the tomographic bin, whereas nDGP produces an enhancement that becomes weaker with increasing $\kappa$ threshold and source redshift. This indicates that the peak 2PCF may be able to help break degeneracy between different MG models. The dependencies on the $\kappa$ threshold have been shown and discussed in Fig.~\ref{fig:P2PCF emu notomo}, and the source-redshift dependence can be explained by similar reasoning. For nDGP, the enhancement is a true effect of the stronger clustering of the density peaks induced by the fifth force, and as such reflects the fifth-force strength which increases over time (i.e., as the redshift decreases). For $f(R)$ gravity, as we explained below Fig.~\ref{fig:P2PCF emu notomo}, the suppression of the peak 2PCF is due to the differences in the peak population, namely the $f(R)$ peak catalogue could include some lower initial density peaks thanks to the enhanced structure growth; in this case, the degree of suppression depends on how many, in relative terms, more peaks are brought into a given catalogue compared to in $\Lambda$CDM, and becomes complicated, but we observe the general trend that the higher the 2PCF is for $\Lambda$CDM the stronger the $f(R)$ suppression tends to be.

\subsection{Forecasts}
\label{sec:forecasts}

In this section we present forecasts for stage-IV surveys for the constraining power of the WL peak abundance and 2PCF on the $f(R)$ and nDGP MG models, whilst simultaneously constraining cosmological parameters. We use stage-IV mock WL maps, cf.~Sec.~\ref{sec:mglens} and Eq.~\eqref{eq: GSN gaussian}, as training data for a GPR emulator in a Bayesian framework coupled with MCMC in order to generate the posteriors presented here (\ref{eq:log likelihood}). In each case we present the peak abundance and 2PCF for the non-tomographic and tomographic cases. First we discuss how these statistics can be used to constrain strong and weak $f(R)$ models, followed by strong and weak nDGP models. This is followed by a test of GR in Sec.~\ref{sec:GR}.

Note that all posteriors presented in this work are deliberately centered on the input 'truth', allowing for easy comparison between posteriors. We note that we have tested and verified that the posteriors are unbiased when not centering on the 'truth'.

\begin{figure*}
    \centering
    \includegraphics[width=2\columnwidth]{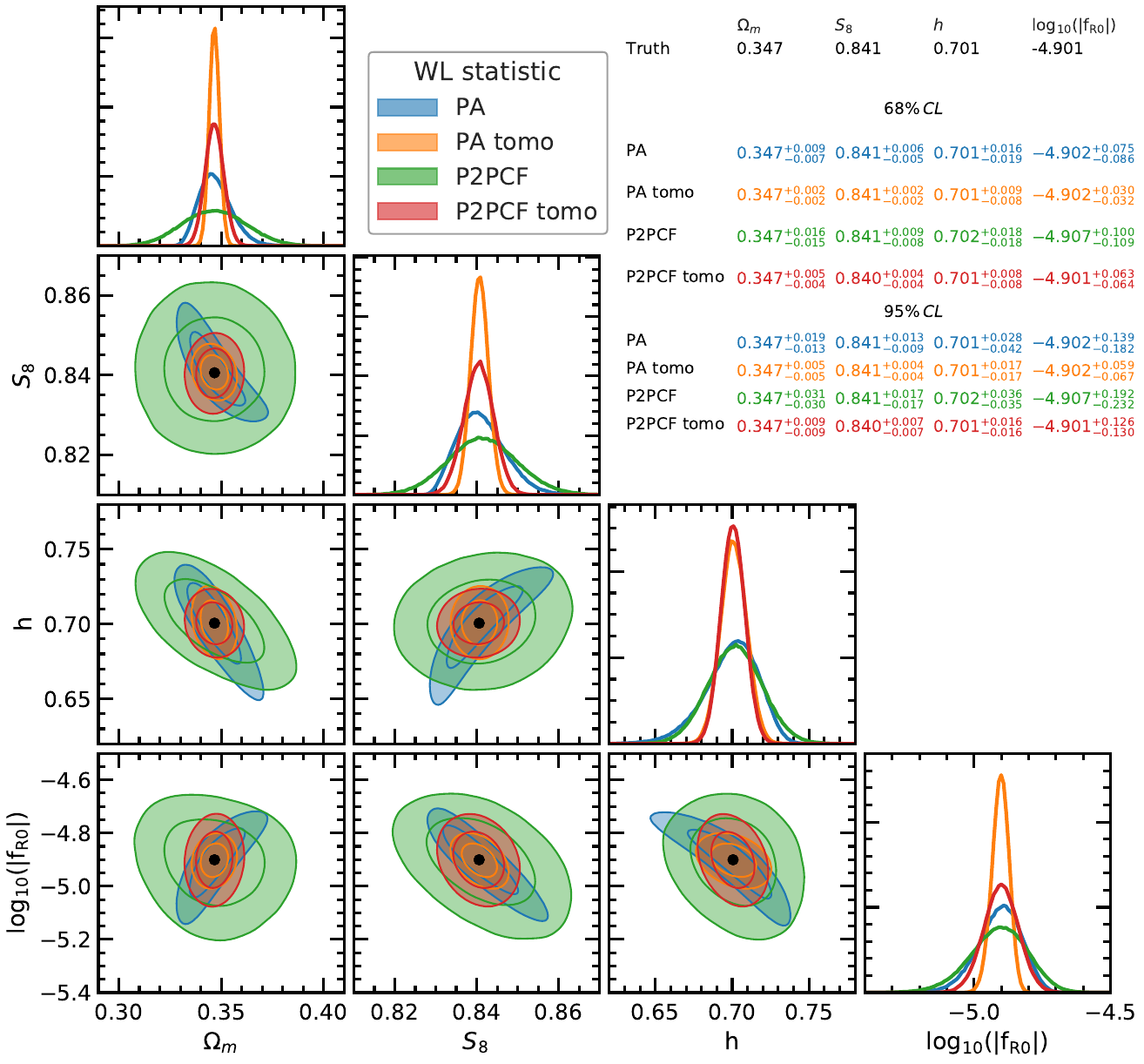}
    \caption{Forecast posteriors for $f(R)$ gravity from a stage IV WL survey. The input mock data corresponds to the cosmological parameter values shown in the first row of the table in the top right. The diagonal panels show the 1D posteriors, and the off-diagonal panels show the 2D posteriors. The orange and blue contours correspond to the constraints from the WL peak abundance (PA) with and without tomography respectively. The red and green contours correspond to the posteriors from the WL peak 2PCF (P2PCF) with and without tomography respectively. The table in the top right quantifies the upper and lower 68\% and 95\% credible levels for each posterior.}
    \label{fig:contours f5}
\end{figure*}

\begin{figure*}
    \centering
    \includegraphics[width=2\columnwidth]{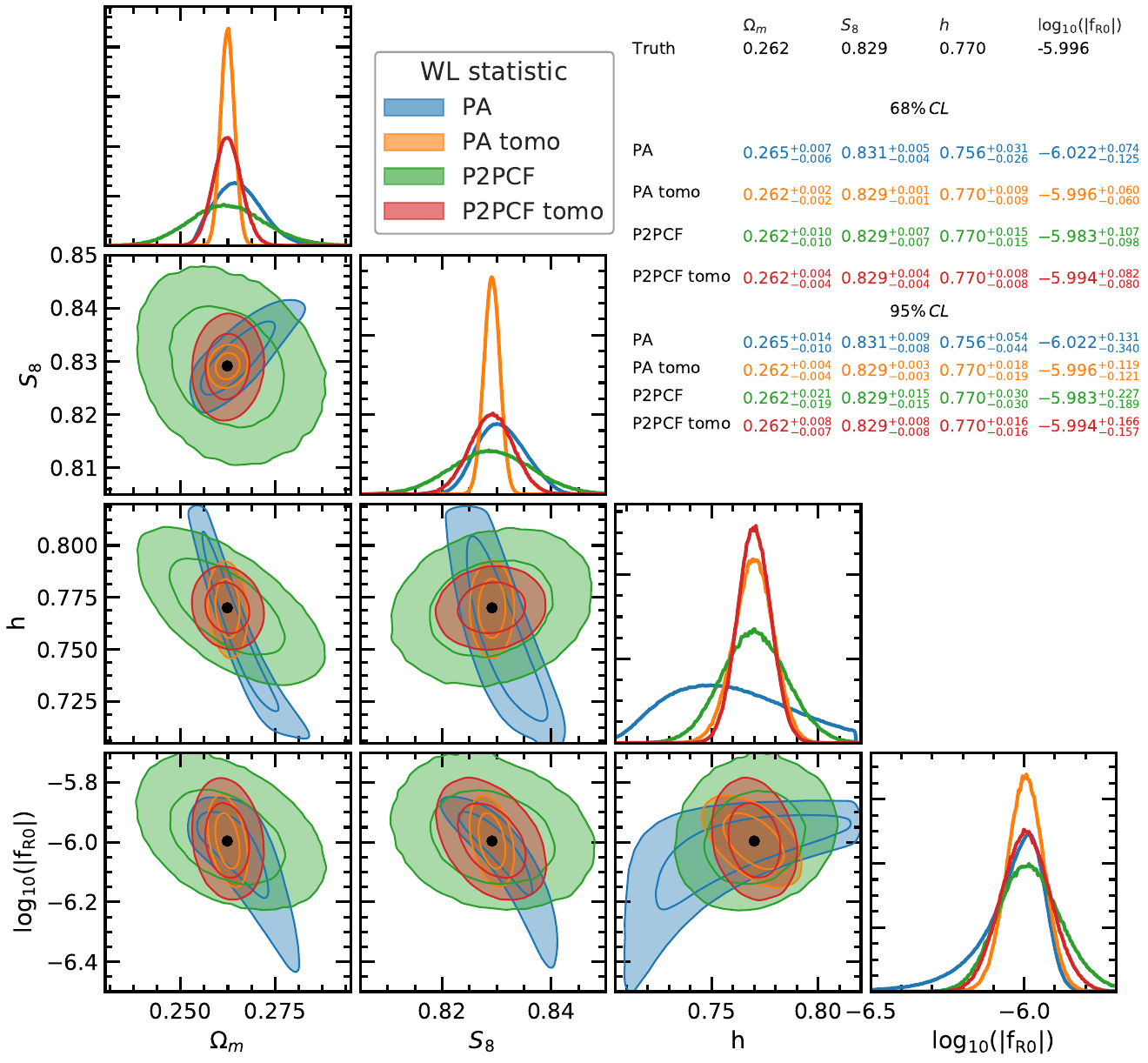}
    \caption{The same as Fig.~\ref{fig:contours f5} except that the mock data is from a weaker $f(R)$ model.}
    \label{fig:contours f6}
\end{figure*}

First we show the strength with which WL peak statistics can be used to constrain our strongest $f(R)$ models, which is presented in Fig.~\ref{fig:contours f5}. The model used as the mock data when generating these posteriors is shown in the table in the top right of the figure, labelled `Truth'. This model has $\log(|f_{R0}|) = -4.9$, where the cosmological parameters depart only slightly from the current concordance cosmology \cite{Planck2018,des/kids:2023}. The constraints from the non-tomographic peak abundance is shown by the blue contours, and the tomographic case is shown in orange. The constraints from the non-tomographic and tomographic peak 2PCF are shown in green and red respectively.

The figure shows that the constraining power of both peak statistics is strongly increased when the tomographic analysis is used. The tomographic analysis increases the strength of the peak abundance constraints by a factor of roughly $2.5$. For the peak 2PCF, the relative increase in constraining power when switching to tomography is slightly lower

In addition, the direction of degeneracy between parameters can be different for the tomographic and non-tomographic peak abundance posteriors, as seen in the $h$-$S_8$ plane. This indicates that the combination of the tomographic and non-tomographic peak abundances would provide even more constraining posteriors, however we do not explore this here. This is similar to the findings reported in \cite{Martinet2020}.

We also note that the direction of degeneracy between different parameters is different, and hence complementary, between the peak abundance and the peak 2PCF. For example, the abundance and 2PCF tomographic contours are orthogonal in the $h$-$S_8$ and $h$-$\log(|f_{R0}|)$ planes. However, the peak 2PCF posteriors completely enclose the abundance posteriors, indicating that the combination of these probes will not benefit substantially from the orthogonal degeneracy of the two peak statistics.

Overall, the peak abundance provides tighter constraints than the peak 2PCF for the strong $f(R)$ model for all parameters except $h$, on which the constraints are very similar for the abundance and 2PCF. The WL peak abundance is able to constrain this model at $2\sigma$ with $\approx 4\%$ precision on the $\log(|f_{R0}|)$ parameter.

For robust tests of MG with WL peaks, GR must be assumed to be the null-hypothesis. It is therefore important to test weaker deviations from GR. In addition, the WL peak abundance and 2PCF posteriors may also change as the MG model becomes weaker, because the degeneracy between the MG and cosmological parameters may depend the strength of the MG model. 

\begin{figure*}
    \centering
    \includegraphics[width=2\columnwidth]{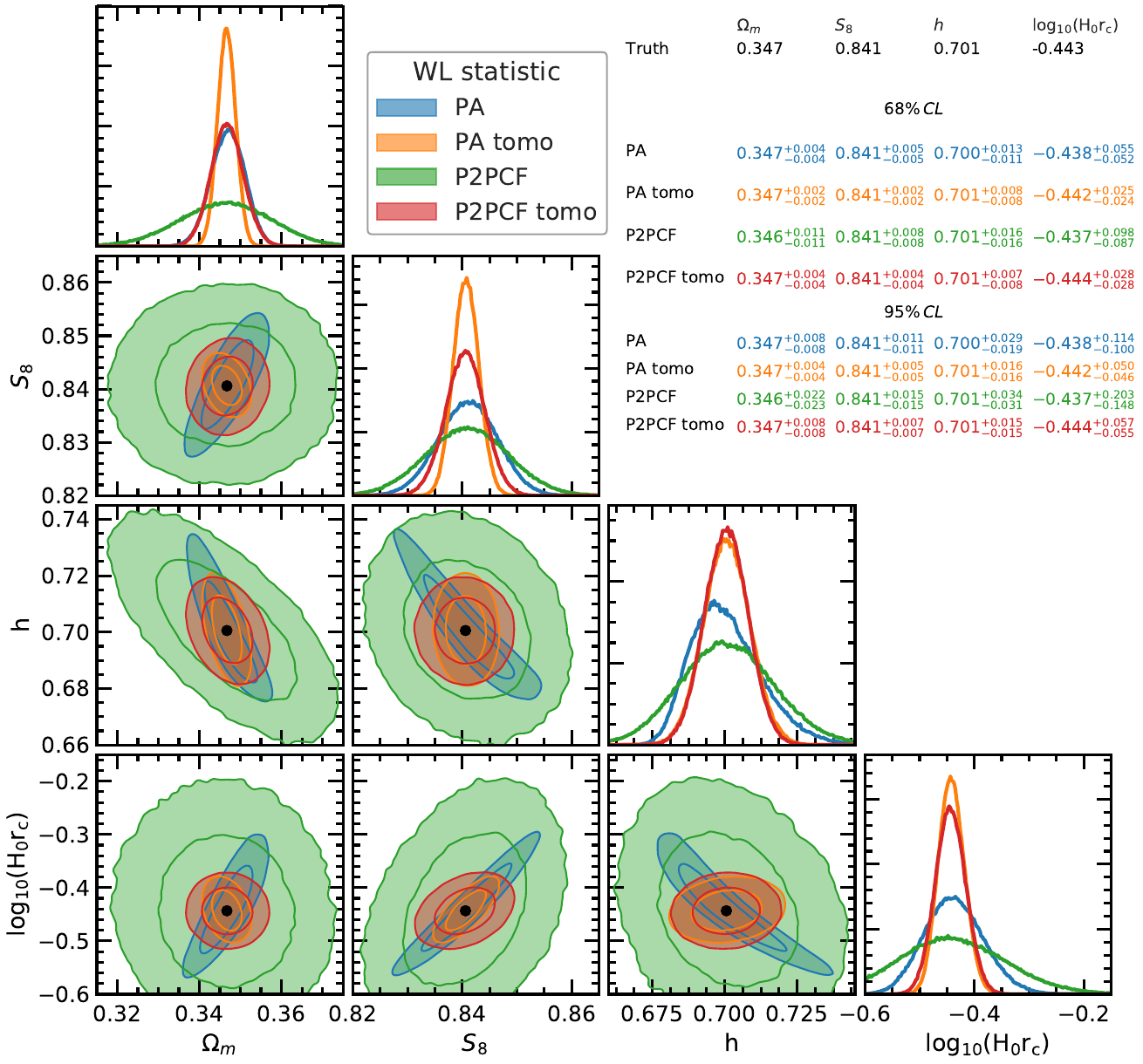}
    \caption{The same as Fig.~\ref{fig:contours f5} except the posteriors shown here are for nDGP gravity.}
    \label{fig:contours nDGP1}
\end{figure*}

\begin{figure*}
    \centering
    \includegraphics[width=2\columnwidth]{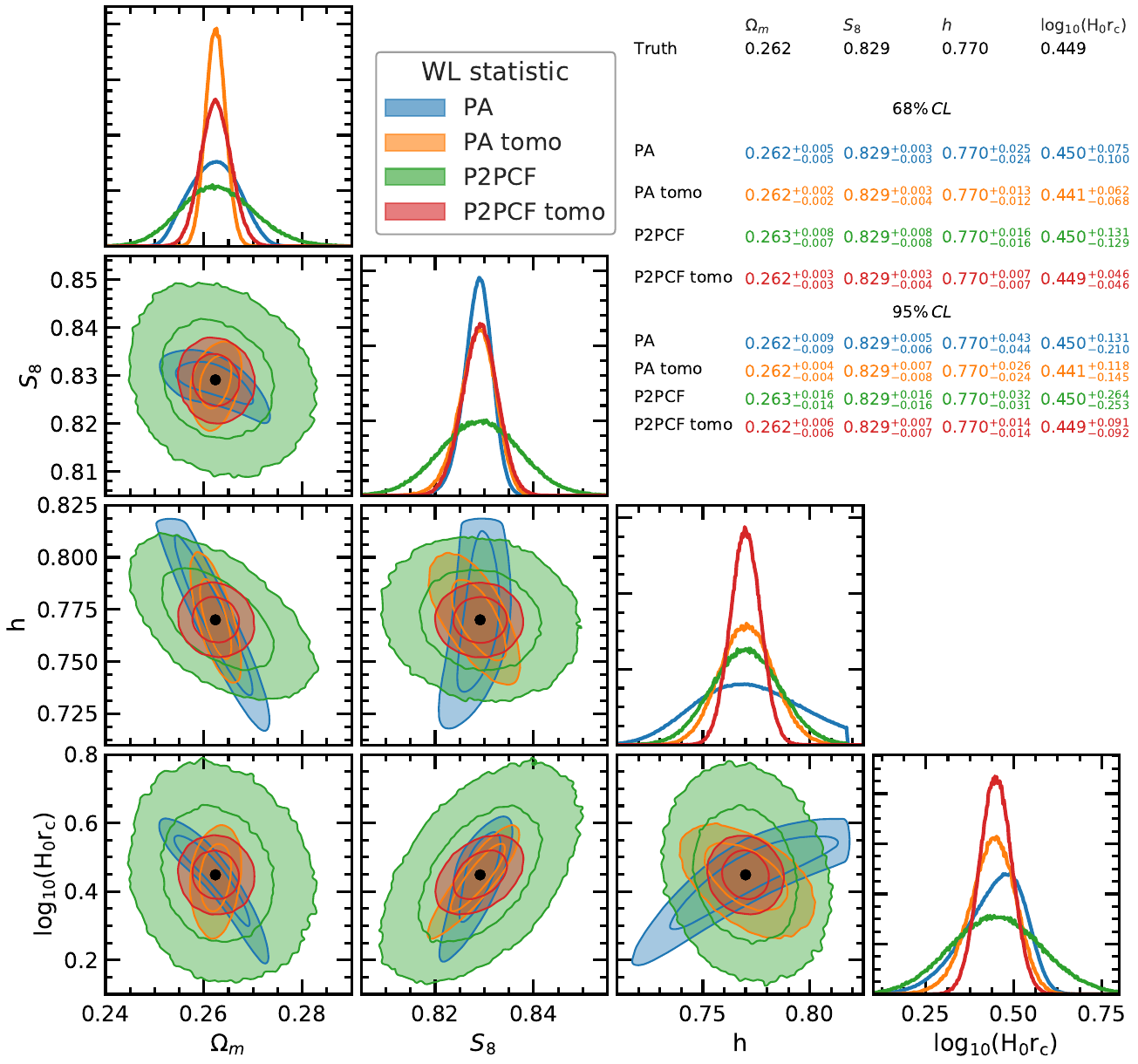}
    \caption{The same as Fig.~\ref{fig:contours nDGP1} except that the `Truth' is a weaker nDGP model.}
    \label{fig:contours nDGP2}
\end{figure*}

Next, in Fig.~\ref{fig:contours f6}, we present the same analysis as before, except now for mock data corresponding to a weaker $f(R)$ model, F6, with $\log(|f_{R0}|) = -5.996$. First, the figure shows that for the non-tomographic case, the peak abundance and peak 2PCF posteriors now provide complementary information to each other, due to different cosmological and MG parameters. This is exemplified by the fact that peak 2PCF contours (green) provide tighter constraints on $h$ than the peak abundance contours (blue) by roughly a factor of two. In addition to this, the peak abundance yields asymmetric constraints on $\log(|f_{R0}|)$, with a larger lower bound uncertainty, which corresponds to poorer constraints in the direction of weaker MG. In contrast, the peak 2PCF provides more symmetric constraints on $\log(|f_{R0}|)$ in this regime. The consequence of this difference is that the abundance and 2PCF constraints also complement each other here, even though the overall magnitude of the uncertainty is very similar.

The figure shows that the tomographic analysis also significantly improves the constraining power of the peak statistics for weak $f(R)$ models, consistent with the constraints for the stronger $f(R)$ model above. When comparing posteriors for the tomographic peak abundance and the tomographic peak 2PCF, the complementarity between the two is diminished relative to the non-tomographic case. This is because, for all parameters except $h$, the tomographic peak abundance gives tighter constraints than the tomographic peak 2PCF, where the former posteriors fully enclose the latter.

In this case, the weak $f(R)$ model with $\log(|f_{R0}|) = -5.996$ can be constrained to $\approx 6\%$ at $2\sigma$ by the tomographic peak abundance. We note that the absolute magnitude of the precision of these constraints is roughly a factor of two weaker than those from the stronger $f(R)$ case above. This is to be expected, because the modified gravity signatures are weaker for the statistics measured for this model (see figures in Sec.~\ref{sec:full n(z)} and \ref{sec:tomo}). 

Next we repeat the same analysis from the first half of this section, but for the nDGP model. Again we first test the precision with which the peak statistics can constrain a stronger nDGP model, followed by case of a weaker nDGP model. 

Fig.~\ref{fig:contours nDGP1} shows constraints on an nDGP model with $\log(H_0r_{\rm c}) = -0.443$, which marks a relatively strong deviation from $\Lambda$CDM, from the same peak statistics. The posteriors presented here follow the same format as the previous figures. Consistent with the $f(R)$ case, the tomographic analysis provides significantly more precision when constraining both the modified gravity and cosmological parameters relative to the non-tomographic case. Overall the tightest constraints on the cosmological parameters come from the tomographic peak abundance, however for $\log(H_0r_{\rm c})$ the precision is comparable for the tomographic peak abundance and tomographic peak 2PCF. The increased sensitivity of the peak 2PCF to nDGP relative to $f(R)$ gravity is consistent with Fig.~\ref{fig:P2PCF emu tomo}, which shows that the magnitude of the modification to the peak 2PCF relative $\Lambda$CDM is larger for the nDGP model. 

When comparing Fig.~\ref{fig:contours nDGP1} to Fig.~\ref{fig:contours f5}, we also see that the directions of degeneracy between cosmological parameters changes for different MG models, as illustrated by the non-tomographic peak abundance in the $S_8$-$\Omega_{\rm m}$ plane. This suggests that there are differences in the degeneracy directions between the MG and cosmological parameters in a given model. Again, this is consistent with the observation that the different MG models modify the peak statistics through different mechanisms.

Next, Fig.~\ref{fig:contours nDGP2} corresponds to the same setup as the previous figure, except now a weaker nDGP model with $\log(H_0r_{\rm c}) = 0.449$ is used as the mock data. In this case we can see that the complementary information between the peak abundance and peak 2PCF is significantly enhanced relative to the stronger nDGP case and the $f(R)$ cases. This behaviour is present for both the non-tomographic and tomographic case, and is due to the fact that the peak abundance provides poorer constraints on $h$ than the peak 2PCF for the weak nDGP model.

\begin{figure*}
    \centering
    \includegraphics[width=2\columnwidth]{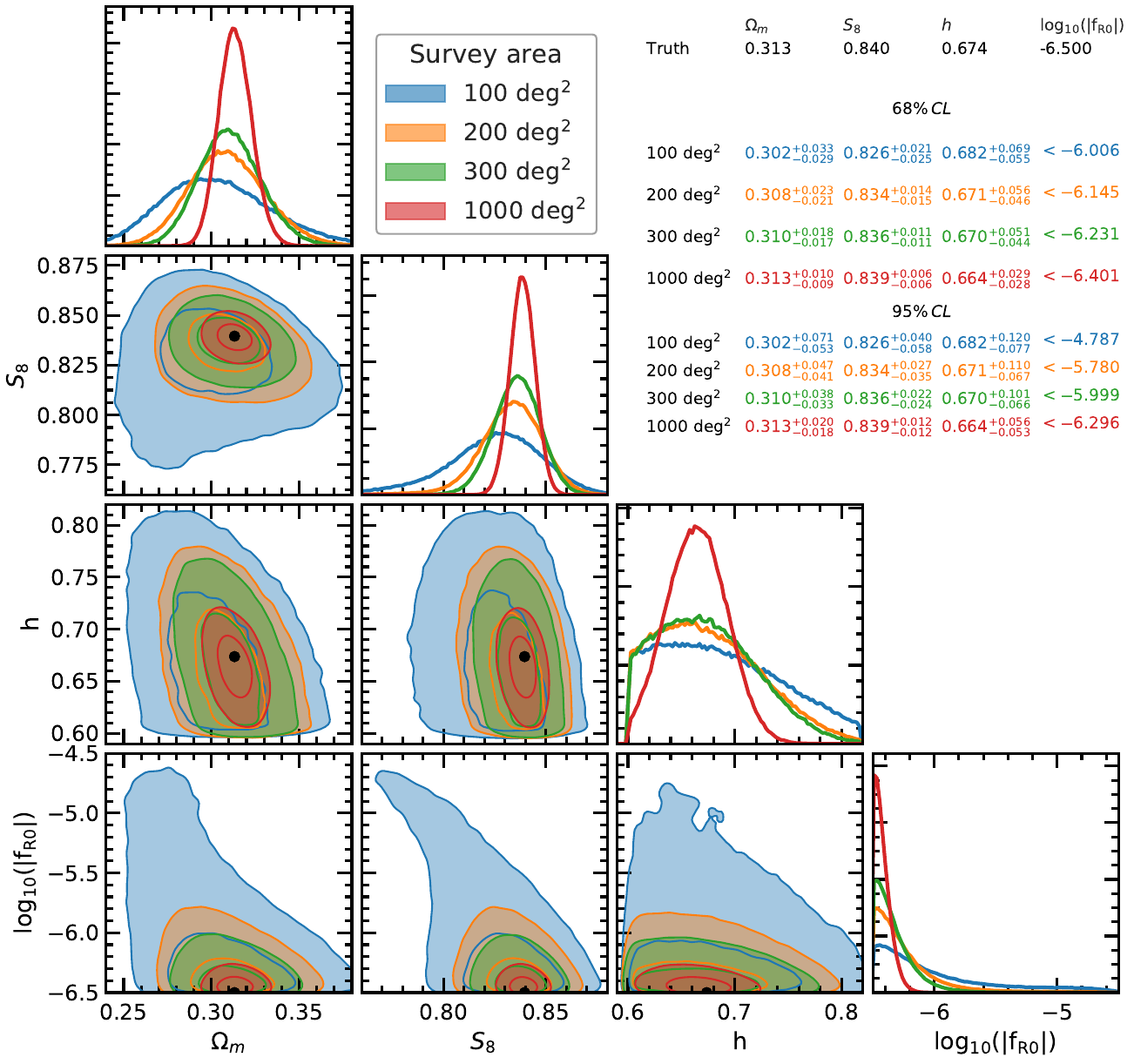}
    \caption{Forecast posteriors for the \textsc{forge} tomographic WL peak abundance. The mock data corresponds to $\Lambda\rm{CDM}$, where the $f(R)$ parameter has been set to $-6.5$ in this case (instead of $-\infty$), in order to effectively train the emulator. The shape noise component of the peak abundance corresponds to that of a stage-IV survey. The survey area for which the covariance is rescaled is varied, leading to the array of posteriors presented as shown by the legend next to the upper-most panel. }
    \label{fig:contours fGR}
\end{figure*}

\begin{figure*}
    \centering
    \includegraphics[width=2\columnwidth]{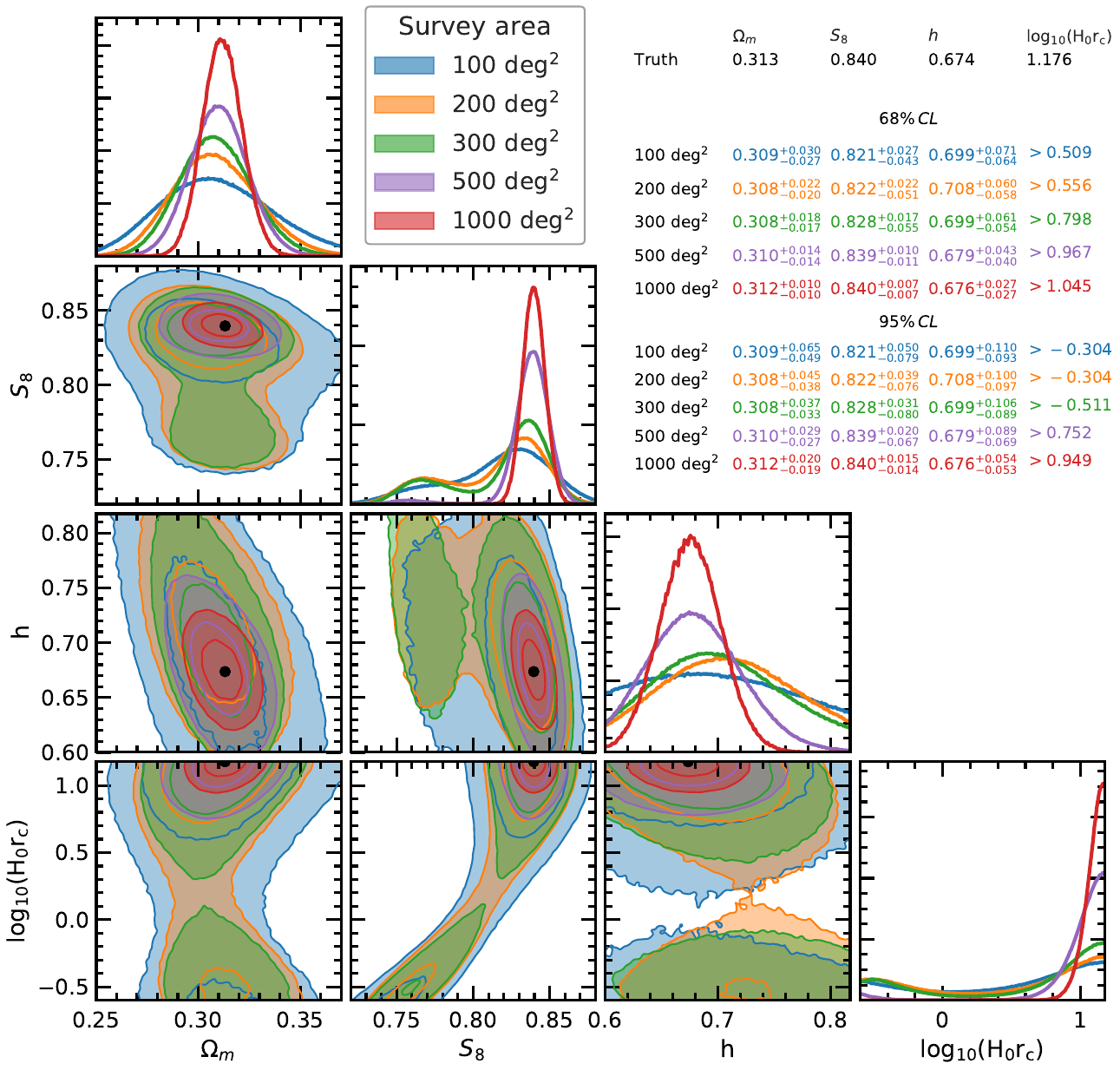}
    \caption{The same as Fig.~\ref{fig:contours fGR} except the \textsc{bridge} tomographic WL peak abundance is used. Here, the nDGP parameter is set to $1.176$ rather than $\infty$ in order to reliably train the emulator.}
    \label{fig:contours nGR}
\end{figure*}

When comparing the tomographic peak abundance to the tomographic peak 2PCF, we see that the constraints on $\Omega_{\rm m}$ are tighter for the peak abundance, and roughly the same for $S_8$. For $h$ and $\log(H_0r_{\rm c})$, however, the precision is significantly greater for the peak 2PCF. This indicates that whilst in most cases studied here the peak abundance outperforms the peak 2PCF, in some cases the 2PCF can be useful for extracting extra information. 

Next we investigate how these statistics can be used to rule out the MG models studied here, which is connected to recovering the input parameters when a GR mock is used.

\subsection{Modified gravity as a test of General Relativity}\label{sec:GR}

One of the main motivations for studying MG models is to test, or constrain possible deviations from, GR. Performing such a test with the pipeline developed in this work comes with additional challenges compared to testing the significance with which MG can be detected. This is because the MG models tend to have a GR limit, i.e., their behaviour goes back to that of GR for extreme parameter values: $-\log|f_{R0}|\rightarrow\infty$ for $f(R)$ gravity and $H_0r_{\rm c}\rightarrow\infty$ for nDGP.

This is challenging for the following two reasons. First, GR exists at the edge of the parameter space over which these models have been sampled, as shown in Fig.~\ref{fig:hypercube}, which means that the resulting accuracy of the emulator at the GR limit is worse relative to the central nodes, where reside the stronger and weaker MG models used previously. This is because the emulator accuracy is greatest in the centre of the parameter space that is sampled, where it is most informed by the surrounding training data. Second, dealing with a limit that corresponds to an infinite value is numerically intractable, which poses an issue for numerical emulation. We deal with these two issues as follows.

First, when considering forecasts for models at the edge of the sampled parameter space, we ask a slightly different question to the one that is studied in Sec.~\ref{sec:forecasts}. Previously, we investigated the precision with which we can constrain strong and weak MG models with a stage IV survey. In this section we instead ask, how much area of a stage IV survey is required in order to rule out the weakest MG model sampled in our training data at the 2$\sigma$ level. This allows us to keep the emulation error at the GR limit small relative to the posterior sizes.

Second, in order to effectively train the emulator down to the GR limit, we replace the infinite values of the MG parameters with values that are slightly outside of the sampled parameter space, essentially tricking the numerical emulator. This has been shown to work in \cite{Harnois-Deraps:2022bie}, where the authors use this approach with the WL convergence power spectrum to constrain the MG parameters with GR mock data. Of course, the resulting accuracy with which the emulator is able to predict the statistics for GR  depends on the numerical value that is chosen to replace the infinite values that correspond to the true GR limit. This is because the total volume of the parameter space will depend on the arbitrary values that are chosen for the GR limit, effectively changing the range over which the emulator must perform its interpolation. If the chosen value is too far from the parameter space where most of the training data exists, then the emulator must perform an interpolation over a large dynamic range, which can lead to compounding errors, and if the chosen value is too close to the edge of the parameter space, the emulator will be more falsely informed about the rate at which the the MG models turn to GR, which will impact the quality of the emulators fit to the rest of the training data. In practice we choose a value that is a compromise between these two competing factors, and quantify the resulting accuracy of the emulator using the cross-validation test, as shown in Appendix \ref{app:emm_acc}.

We perform this test for both of MG emulators, as the exact nature in which we probe GR depends on the MG model considered. 

Fig.~\ref{fig:contours fGR} shows posteriors from the peak abundance for a range of stage-IV survey areas -- $100, 200, 300$ and $1000$ $\rm{deg}^2$, with the corresponding colours indicated in the legend. We set the ``GR value'' of $\log(|f_{R0}|)$ to $-6.5$, which allows us to emulate between $f(R)$ gravity and GR within a finite parameter volume. The posteriors show that the constraints on the cosmological parameters and the MG parameter increase with survey area, as expected.

Interestingly, we find that $\log(|f_{R0}|)$ can be constrained to $<-5.999$ at the $2\sigma$ level with only $300$ deg$^2$ of stage-IV survey area. The gain in precision with which the MG parameter can be constrained as the survey area increases further gives diminishing returns. This is somewhat expected due to the nature of the $f(R)$ parameter, where $\log(|f_{R0}|) = -5$ corresponds to MG ten times stronger than $\log(|f_{R0}|) = -6$, indicating that it becomes exponentially more difficult to rule out weaker and weaker models.

Fig.~\ref{fig:contours nGR} is the same as Fig.~\ref{fig:contours fGR} but for the case of nDGP, for which we have set the value of $\log(H_0r_{\rm c})$ to $1.176$ for the ``GR node''. The figure shows that for small survey areas, there is a strong degeneracy between $S_8$ and $\log(H_0r_{\rm c})$ for GR mock data We have added an additional contour corresponding to 500 $\rm{deg}^2$ of a stage IV survey, in order to highlight the regime in which this degeneracy is no longer apparent. The weakest nDGP model that is simulated by \textsc{bridge} has $\log(H_0r_{\rm c}) = 1.0$, therefore the posteriors indicate that we can rule out similar models at the 2$\sigma$ level with 1000 deg$^2$ of survey area.

\section{Discussion and Conclusions} \label{sec:conclusion}

In this paper, we measure and present WL peak statistics from the \textsc{forge} and \textsc{bridge} \citep{Arnold:2022,Harnois-Deraps:2022bie} simulations, and their resulting WL maps \citep{Harnois-Deraps:2022bie}, which are made to match the specifications of a stage-IV survey with five tomographic bins, as shown in Fig.~\ref{fig:n(z)}. We then perform analyses of the constraining power of two WL peak statistics, the peak abundance and 2PCF, for both non-tomographic and tomographic WL convergence maps.

By using the GPR emulator to investigate the response of the peak statistics to modified gravity, we find that in the non-tomographic case, both MG models enhance the abundance of high amplitude peaks, as well as increasing the abundance of peaks with an amplitude with $\kappa < 0$. Quantitatively, the nDGP model leads to a greater increase in the number of negative peaks relative to $f(R)$ gravity, as well as an increase in the abundance of high kappa peaks over a broaded $\kappa$ range. In contrast, the highest $\kappa$ peaks in $f(R)$ gravity are broadly unaffected relative to $\Lambda$CDM, which is due to the different screening mechanisms present in the $f(R)$ and nDGP models. For $f(R)$ gravity, the chameleon mechanism yields a short range fifth force, which predominantly enhances structure formation in lower mass haloes, whereas for nDGP, the Vainshtein mechanism yields a long range fifth force, which is more efficient at enhancing the formation of high mass haloes.

For the non-tomographic WL peak 2PCF, we find that the effect of MG goes in opposite directions for the $f(R)$ and nDGP models. In the $f(R)$ case, the clustering of WL peaks is suppressed, and this suppression becomes stronger as increasingly-higher $\kappa$ peaks are removed from the 2PCF calculation. In contrast, the peak clustering is enhanced in nDGP, and this enhancement becomes weaker as increasingly-higher $\kappa$ peaks are removed from the 2PCF measurement. This difference is again due to the different screening mechanisms in the two MG models.

In nDGP, which features a long range fifth force, the clustering of the haloes is directly enhanced, and this dominates the contribution to the change in the 2PCF compared to the mass enhancement of less clustered haloes. This leads to an enhancement of clustering, which decreases as the kappa threshold increases. The reason this enhancement decreases with increasing kappa threshold is because the long range fifth force is screened for the highest mass haloes.

For the tomographic peak abundance, the responses to the two MG models are qualitatively similar to the respective non-tomographic cases, but the effect is weaker for the low redshift tomographic bins, increasing to higher redshift bins. It is also apparent that the nDGP model yields a stronger enhancement of the peak abundance when compared to $f(R)$. This is because nDGP increases the power of larger scale modes compared to $f(R)$, which are known to contribute to the peak abundance \citep{Sabyr2022}. 

For the tomographic peak 2PCF, the results are qualitatively similar to the non-tomographic case. For $f(R)$, the suppression of the 2PCF is stronger for higher source redshifts, which is due to the lower relative noise contribution to the physical signal. Where the noise is suppressing the small scale features induced by $f(R)$ gravity. For nDGP, the lower source redshifts produce a stronger enhancement compared to high source redshifts, this is because the large scale modes are less impacted by noise (which is suppressed on large scales via smoothing). Therefore the impact of MG, which is more dominant at lower redshifts, is more apparent. 

Finally, we have shown how the peak statistics studied in this work can be used to constrain $f(R)$ and nDGP models, as well as the survey area  required to rule out weak MG models in our simulation suites at the $2\sigma$ level. We found that tomographic analysis significantly increases the constraining power of both WL peak statistics, and that the peak 2PCF provides complementary information to the peak abundance for the weakest MG models. For $f(R)$ gravity the tomographic peak abundance can constrain strong and weak models at $\log(f_{R0}) = -4.9\pm0.03$ and $-6.0\pm0.06$ respectively. For nDGP we find $\log(H_0r_{\rm c}) = -0.4\pm0.03$ and $0.44\pm0.07$ for strong and weak models respectively. In the case of using these MG emulators to test GR, we use GR mock data and vary the effective survey area. This allows us to find the survey area required to rule out the weakest MG models in \textsc{forge} and \textsc{bridge} at the two sigma level. We find that roughly $1000 \rm{deg}^2$ of stage-IV survey area is required. We note that these forecast constraints assume that it will be possible to reliably exploit information from the convergence field smoothed over a one arc-minute scale.  

The forecasts presented in the previous sections illustrate the power of WL peaks in constraining MG models. These forecasts employ a noise level and $n(z)$ that is expected from a stage-IV survey, and show that such a dataset may yield constraints on MG that improve upon current constraints from cosmological scales \citep[e.g.][]{X.Liu2016, artis2024, Vogt2024b}. However, systematics such as baryonic physics, source clustering, intrinsic alignments, and photo-$z$ uncertainties have not been included in this analysis. These forecasts therefore assume that these systematics can be perfectly understood when stage-IV data arrives. In practice this is too optimistic, and these systematics must be accounted for through forward modelling in future works. This work therefore illustrates the maximum potential constraining power of WL peaks for MG cosmologies.

The cosmological constraining power that the WL peak statistics exhibit in this work is driven by the small scales, owing to the fact that the WL maps are smoothed with a one arcminute Gaussian filter. It will therefore be crucial to reliably model baryonic physics and galaxy intrinsic alignments in this regime when these statistics are extracted from real data, as these are the two systematic uncertainties that dominate the small scales. Modelling these systematics is extremely challenging, and the most accurate way to achieve this is through hydrodynamical simulations.

Recently, state-of-the-art hydrodynamical simulations such as \textsc{MilleniumTNG} \citep{Pakmor_2023} and \textsc{FLAMINGO} \citep{Schaye_2023} have been used to construct realistic mock WL maps that include baryons \citep{Ferlito_2023, Broxterman_2024}. The authors use these WL maps to study the impact of baryons on the WL peak abundance, and find that this leads to an overall suppression of the WL peak abundance, where \cite{Ferlito_2023} demonstrated that this suppression increases with increasing $\kappa$ and decreasing source redshift. Given that the scatter in the WL peak abundance increases with increasing $\kappa$ (where the baryonic suppression dominates), the cosmological constraining power from the WL peak abundance is driven by the region near the peak of the distribution, which is roughly between $\kappa = 0$ and $\kappa = 0.05$, where the baryonic suppression is less severe. Over the $\kappa$ ranges used in this analysis, we expect tomographic bin one to be the most affected, by roughly $2\%$ to $10\%$, and for tomographic bin five to be the least affected by roughly $1\%$ to $5\%$ \citep{Ferlito_2023}. Therefore, a robust treatment of baryonic physics when applying the pipeline presented in this work to real data would require the inclusion of the baryonic suppression. This could be modelled by taking the ratio of the dark matter only peak abundance to the hydro peak abundance \citep[e.g those presented in][]{Ferlito_2023, Broxterman_2024} and including it as a multiplicative factor to the model peak abundances presented in this work (we note that such an approach has been confirmed to work for the power spectrum \citep{Arnold:2019} and for the halo mass function \citep{elbers2024}). This approach would therefore not alter the posterior distributions shown here (as the rescaling would be the same for all models), and would allow for unbiased constraints when applied to real data. This approach relies on the assumption that the \textsc{MilleniumTNG} and \textsc{FLAMINGO} peak abundances are consistent and do not depend on the underlying subgrid models, and that they both accurately represent baryonic suppression in the real Universe, which remains to be tested. Therefore, it will be crucial to also vary the sub-grid parameters when evaluating the baryonic suppression factor, following the approach in \cite{Salcido_2023}, which will add additional free parameters to the analysis. In addition to this it will also be important to consider the interactions between modified gravity and baryonic physics, as studied in \cite{Mitchell_2022}, when constructing the dark matter to hydro peak abundance ratio. However, at present there are no WL maps from these simulations, and so we leave such studies to a future work.

The intrinsic alignments of galaxies also play a key role in the WL signature at small scales, which leads to significant contamination if not modelled correctly. A range of theoretical models for galaxy intrinsic alignments exist \citep{Lamman_2024}, that can be used to create a forward model which produces WL maps that include contributions from intrinsic alignments \citep{harnois2021}. These models introduce additional free parameters which represent the strength of the intrinsic alignment, which should be varied alongside the cosmological parameters as in \cite{Zurcher2022}. Additionally the intrinsic alignment signal of galaxies has been simulated in a full hydrodynamical setting in \cite{Delgado_2023} which shows that it may be possible to include simulated intrinsic alignments in WL maps in the future. The additional intrinsic alignment parameters will lead to a broadening of the posteriors shown in this work, but we leave the interaction between intrinsic alignments and modified gravity to a future study.

Lastly, weak lensing data are inevitably affected by residual uncertainty on the redshift distribution of the photometric source galaxies and on the shape measurement calibration, two effects that need to be forward-modelled at the level of ray-tracing in a full data analysis, as in e.g. \cite{Zurcher2022}. We note that neglecting to model the photometric error will lead to inconsistencies between the simulated source redshift distribution and the true distribution, and will induce biases when applied to real data, however we do not expect it to significantly alter the conclusions of this work.

Whilst modelling the above-mentioned additional systematic uncertainties may reduce the constraining power of the WL peaks presented here, there are also a range of options to further increase the WL peak constraining power. One approach is to measure the peak statistics for a range of smoothing scales \citep[e.g.]{J.Liu2015,Kacprzak2016,Zurcher2022}. Additionally, the tomographic analysis used in this work, whilst shown to increase the power of the constraints relative to no tomography, can be pushed even further into higher order combinations of tomographic bins, as outlined in \cite{Martinet2020}.

Finally, the covariance matrix used in this analysis is evaluated using the fiducial $\Lambda$CDM cosmology. Typically, when a covariance is assumed to be constant as a function of cosmological parameters, it is evaluated at the center of the parameter space that is sampled. In our case, however, the covariance matrix is effectively at the edge of the parameter space. Where the resulting posterior indicates a detection of modified gravity, this would likely induce biases.
However, given that when testing modified gravity, $\Lambda$CDM corresponds to the null hypothesis, it is reasonable to use a $\Lambda$CDM+GR covariance matrix if the cosmological parameters cannot be varied when estimating the covariance.

\section*{Acknowledgements}
CTD acknowledges support from the Ludwig Maximilians-Universität in Munich. JHD acknowledges support from an STFC Ernest Rutherford Fellowship (project reference ST/S004858/1). 
BL acknowledges support from STFC Consolidated Grants ST/I00162X/1, ST/P000541/1 and ST/X001075/1.
BG acknowledges support from the UKRI Stephen Hawking Fellowship (grant reference EP/Y017137/1) and a UK STFC Consolidated Grant.
CH-A acknowledges support from the Excellence Cluster ORIGINS which is funded by the Deutsche Forschungsgemeinschaft (DFG, German Research Foundation) under Germany's Excellence Strategy -- EXC-2094 -- 390783311. 

The simulations of this project made use of the DiRAC@Durham facility managed by the Institute for Computational Cosmology on behalf of the STFC DiRAC HPC Facility (\href{www.dirac.ac.uk}{www.dirac.ac.uk}). The equipment was funded by BEIS capital funding via STFC capital grants ST/K00042X/1, ST/P002293/1, ST/R002371/1 and ST/S002502/1, Durham University and STFC operations grant ST/R000832/1. DiRAC is part of the National e-Infrastructure.




\bibliographystyle{mnras}
\bibliography{bib} 



\appendix


\begin{figure*}
    \centering
    \begin{subfigure}{\columnwidth}
        \includegraphics[width=\columnwidth]{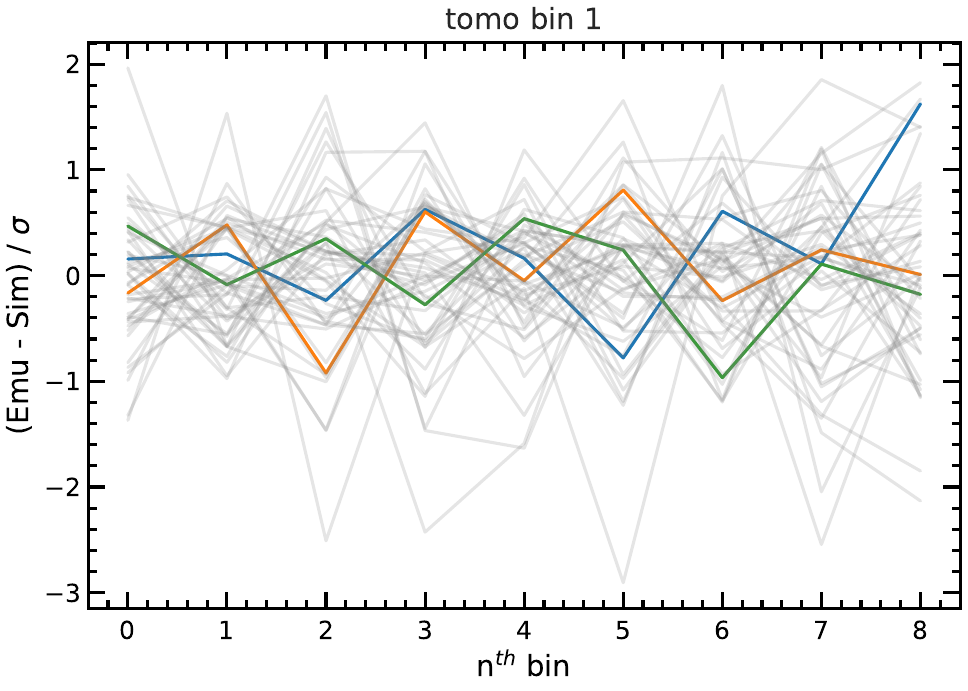}
    \end{subfigure}
    \begin{subfigure}{\columnwidth}
        \includegraphics[width=\columnwidth]{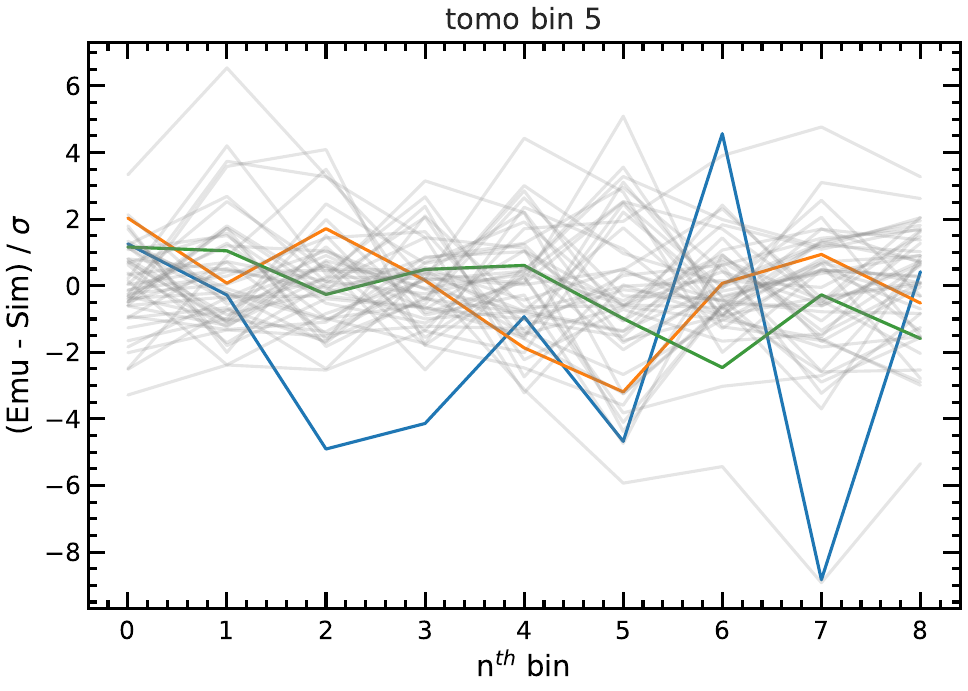}
    \end{subfigure}
    \caption{The cross validation test for the \textsc{forge} WL peak abundance emulator. Each of the 50 curves corresponds to the accuracy with which the peak abundance is emulated at a given node, when that node is removed from the training data. The accuracy of the emulator is quantified as the difference between the simulated peak abundance and the emulated peak abundance, relative to the standard error of the simulated peak abundance. The $x$-axis indicates the n$^{\rm th}$ bin of the emulated statistics. The left and right panels show the cross validation tests for the $1^{\rm{st}}$ and $5^{\rm{th}}$ tomographic bins respectively. The blue, orange, and green curves show the accuracy tests at the nodes corresponding to the mock data used in Figs.~\ref{fig:contours f5}, \ref{fig:contours f6}, and \ref{fig:contours fGR} respectively. The left panel (tomographic bin 1) corresponds to the best case accuracy of the emulator, and the right panel (tomographic bin 5) corresponds to the worst case. Therefore we only show the two most extreme cases for brevity.}
    \label{fig:PA cross val}
\end{figure*}

\begin{figure}
    \centering
    \includegraphics[width=\columnwidth]{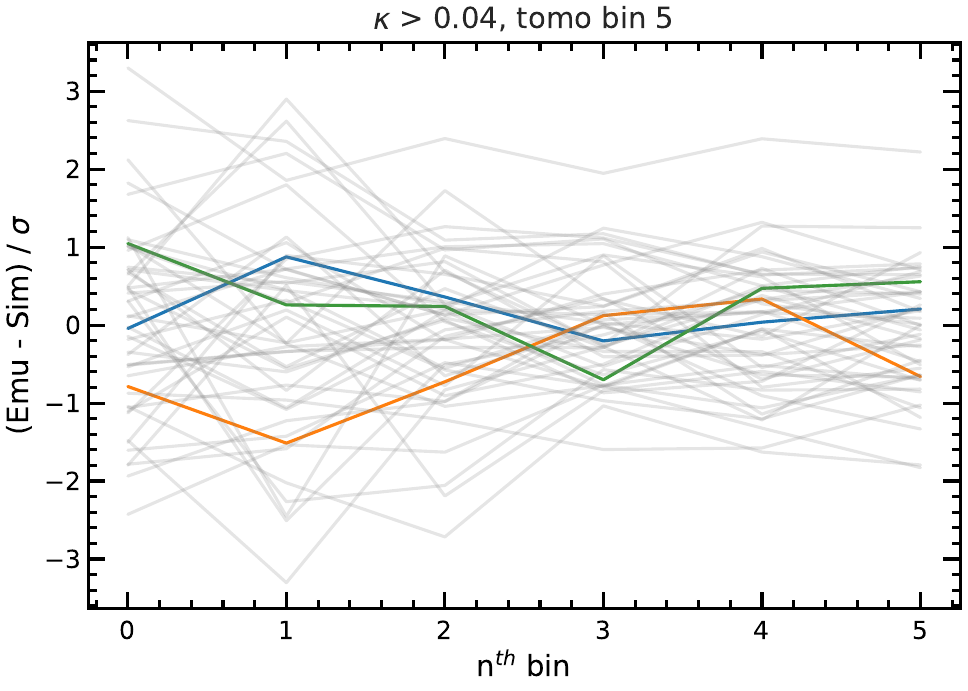}
    \caption{The same as Fig.~\ref{fig:PA cross val} but for the WL P2PCF in the $5^{\rm{th}}$ tomographic bin, with $\kappa > 0.04$. The accuracy of the WL P2PCF emulator for other tomographic bins and $\kappa$ thresholds is similar to that presented here, and so we only show a single case for brevity.}
    \label{fig:P2PCF cross val}
\end{figure}

\begin{figure*}
    \centering
    \begin{subfigure}{\columnwidth}
        \includegraphics[width=\columnwidth]{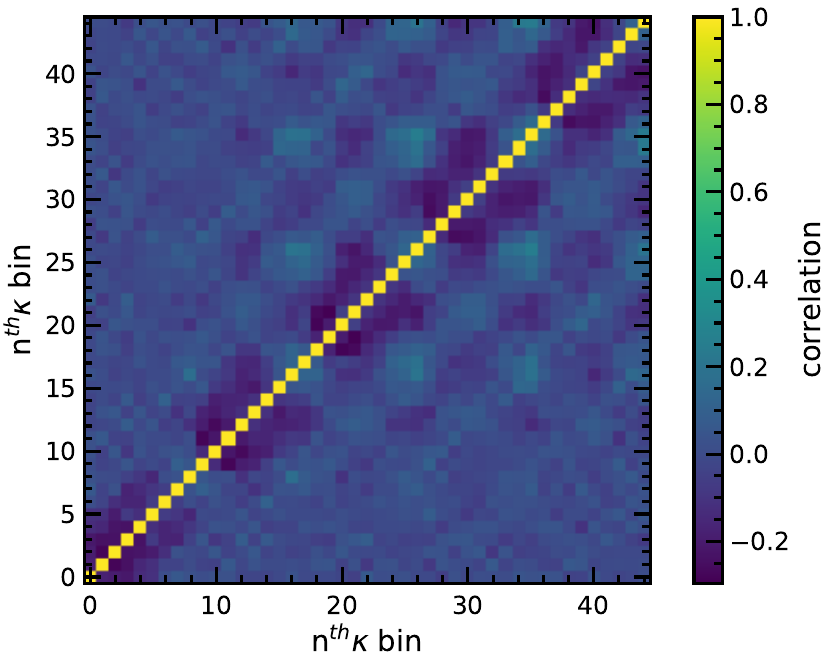}
    \end{subfigure}
    \begin{subfigure}{\columnwidth}
        \includegraphics[width=\columnwidth]{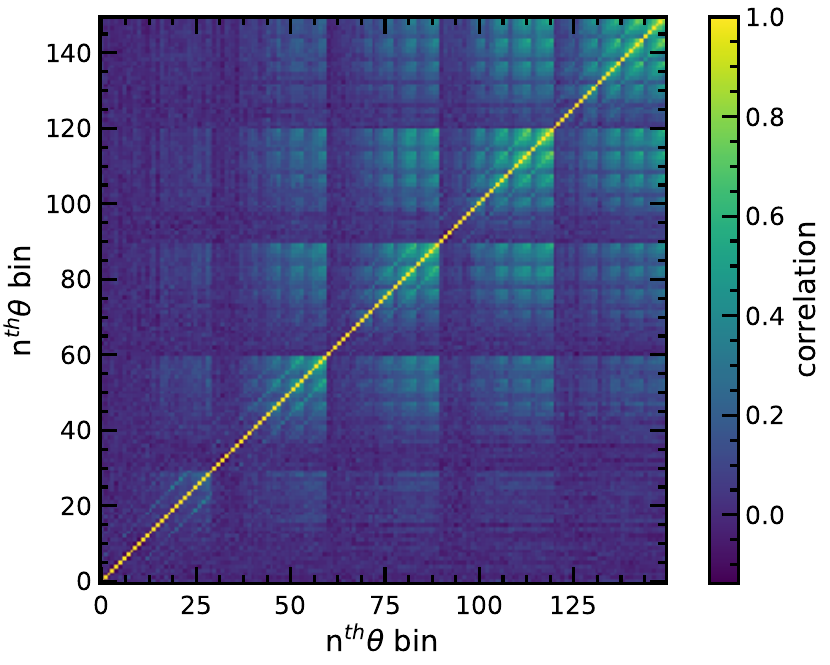}
    \end{subfigure}
    \caption{Left: The joint correlation matrix of the WL peak abundance for the five tomographic bins studied in this work. Right: The joint correlation matrix of the WL P2PCF for all $\kappa$ thresholds and tomographic bins studied in this work. The colour bars to the right of the sub-panels indicate the relative correlation level between two bins in the 2D matrices. Note that the covariance and corresponding correlation matrices are evaluated at a $\Lambda\rm{CDM}$ cosmology. }
    \label{fig:corr}
\end{figure*}

\section{Emulation accuracy}\label{app:emm_acc}
In this appendix we present the cross-validation test performed on the GPR emulator. This allows us to quantify the accuracy with which the emulator is able to predict the WL statistics. The cross-validation test is outlined as follows. One node is removed from the training sample and used as a test sample. The data from the remaining 49 nodes are used to train emulator, and then we use the emulator to predict the peak statistics for the parameters of the test sample. We then compute the difference between the emulated and simulated test samples, and divide it by the standard error on the simulated test sample. We repeat this procedure 49 more times, once for each node, by moving it from the training set to the test set. This gives an indication of how well the emulator is able to predict the peak statistics for data on which it is not trained, within the parameter space that is sampled. Given that the emulator used in this work is trained on all 50 nodes, we can expect that the cross-validation test slightly under estimates the accuracy of the final emulator. The results of the cross-validation test for the peak abundance are shown in Fig.~\ref{fig:PA cross val}, and for the peak 2PCF in \ref{fig:P2PCF cross val}. We note that the panels presented in Fig.~\ref{fig:PA cross val} correspond to the best and worse case scenarios of the emulators performance, and the results shown in Fig.~\ref{fig:P2PCF cross val} are representative of the emulators accuracy for other peak catalogues and tomographic bins. The results presented here are for the \textsc{forge} suite, and we find consistent results with \textsc{bridge}, which we do not show for brevity.

As shown in the figures, for the $\Lambda$CDM cosmology (blue), the accuracy is poorer than that for the strong and weak MG models (orange and green respectively). This is because the fiducial model lies at the boundary of the parameter space that is sampled, where the emulator is least informed by the training data. The poorer accuracy of the emulator in this regime is part of the motivation behind how we test the power with which GR can be recovered with the MG emulators in Sec.~\ref{sec:GR}, which is to find the survey area required to rule out the weakest MG models that are sampled in the simulation suites. 

Overall we find that the cross-validation tests show that the emulators are sufficiently accurate, as they are able reproduce simulated data that they are not trained on within one standard error for the majority of cases. However, for the high-redshift case the accuracy is slightly lower, where the accuracy degrades with increasing redshift. This is because for fixed noise levels per tomographic bin, the higher redshifts have higher convergence with signal-to-noise, which also means more variations with respect to cosmology and gravity. The increased sensitivity to the underlying complex physical model means that these features are harder to emulate.

\section{covariance matrices}\label{app:cov_mat}

In this appendix we present correlation matrices, which are rescaled versions of the covariances matrices used in the generation of the posteriors shown in this work. Fig.~\ref{fig:corr} shows the correlation matrices of the tomographic peak abundances (left) and the tomographic peak 2PCFs (right). The peak abundance matrix shows a small degree of anti-correlation between the low-$\kappa$ and the high-$\kappa$ peak counts in a given tomographic bin, with a similar but weaker patter across tomographic bins. For the peak 2PCFs, the correlation matrix shows that separate bins for a given peak catalogue and tomographic bin are highly correlated, which is expected as the 2PCFs appear to follow a power-law. Additionally, the correlation of the 2PCFs between different catalogues is very strong, which is due to the fact that many of the peaks in both of these catalogues are the same. By capturing this correlation in the covariance matrix, we ensure that we are not including duplicate information in the derived posteriors. Finally, as with the peak abundance, nearby tomographic bins are also correlated for the 2PCFs, which is because of the overlapping redshift ranges of the foreground matter for the different source redshift bins. 


\bsp	
\label{lastpage}
\end{document}